\documentclass[pra,reprint,superscriptaddress,amsmath,amssymb,floatfix]{revtex4-2}

\usepackage{bm}
\usepackage{graphicx}
\usepackage{mathtools}
\usepackage{physics}
\usepackage{microtype}
\usepackage{xcolor}
\usepackage{hyperref}
\usepackage{orcidlink}

\hypersetup{colorlinks=true,linkcolor=blue,citecolor=blue,urlcolor=blue}

\newcommand{\I}{\mathbb{I}}
\newcommand{\Ag}{\mathbf A_g}
\newcommand{\Cf}{\mathcal C_f}
\newcommand{\Df}{\mathcal D_f}
\begin{document}

\title{Revisiting structured-light propagation in screw-dislocation geometries: Landau dynamics and gradient-induced polarization textures}

\author{Edilberto O. Silva\,\orcidlink{0000-0002-0297-5747}}
\email{edilberto.silva@ufma.br}
\affiliation{Programa de P\'os-Gradua\c{c}\~ao em F\'isica \& Coordena\c{c}\~ao do Curso de F\'{\i}sica -- Bacharelado, Universidade Federal do Maranh\~ao, S\~ao Lu\'{\i}s, Maranh\~ao 65080-805, Brazil}

\date{\today}

\begin{abstract}
We derive the propagation of finite electromagnetic beams in the effective metric of a continuous distribution of screw dislocations from gauge-invariant, minimally coupled Maxwell theory with $F=dA$.  The Burgers flux fixes the metric profile through $\sigma(\rho)=f'(\rho)/(b\rho)$, which separates true density gradients from coordinate factors.  In geometric optics, both polarizations share one light cone, and a uniform density rotates the polarization uniformly relative to the geometry-adapted frame; laboratory polarimetry also requires matching at the interfaces.  At leading paraxial order, the orbital motion is of Landau type, with a closed propagator for arbitrary inputs, and for a uniform density the polarization evolution factorizes exactly.  For an arbitrary smooth axisymmetric density, the polarization texture initially grows in proportion to the standard deviation of the density weighted by the beam intensity; the local-phase coherence is the characteristic function of that distribution.  For a coherent transverse Jones field, an exact relation links the linear-Stokes texture and the local circular polarization to spatial--polarization nonseparability.  For linearly polarized, real axisymmetric input beams, diffraction produces local circular polarization at second order in the propagation distance, while the integrated circular Stokes component is conserved in the lossless leading-paraxial theory.  Exact Maxwell eigenmodes, analytic results for a saturating profile, and numerical propagation test the approximation.  We propose a full-Stokes measurement that compares uniform and graded samples with the same mean twist, with a closed-form criterion for the beam waist.  These predictions apply to the ideal metric-Maxwell model or to a transformation-optics medium that realizes it, not to natural dislocated crystals without further constitutive modeling.
\end{abstract}

\maketitle

\section{Introduction}

Structured light exploits spatially varying amplitude, phase, polarization, and angular momentum to encode and manipulate optical degrees of freedom beyond the plane-wave limit~\cite{Rubinsztein2017,Forbes2021}.  In parallel, the geometric theory of defects provides a continuum description in which dislocations are associated with torsion and disclinations with curvature~\cite{Bilby1955,Kroner1981,Katanaev1992,Lazar2010}.  For a continuous distribution of parallel screw dislocations, Silva Netto and Furtado studied the homogeneous metric that underlies the uniform limit considered here and showed its Landau-like quantum dynamics~\cite{SilvaNetto2008}.  The screw-dislocation geometry is thus a convenient place to ask how a finite electromagnetic beam propagates when the propagation equations follow from the metric alone, with no separately postulated optical coupling.

Photonic gauge fields and Landau quantization are by now established themes in structured photonics, with synthetic magnetic fields realized in twisted resonators~\cite{Schine2016}, predicted for strained photonic crystals~\cite{Guglielmon2021}, and observed directly as photonic Landau levels in strained photonic-crystal slabs~\cite{Barczyk2024,Barsukova2024}; see Ref.~\cite{Song2025} for a review.  Likewise, mechanically twisted fibers and waveguides are known to exhibit optical activity, circular and orbital-angular-momentum (OAM) birefringence, and frame-dependent modal effects~\cite{Xi2013,Barshak2015,Buerger2024}.  These systems are useful points of comparison, but in each of them the effect comes from engineered resonator phases, photonic-crystal strain, mechanical twist, stress, or rotating microstructure.  Here we ask a narrower question: what follows when the only electromagnetic dynamics is minimally coupled Maxwell theory on the effective metric of a continuum of screw dislocations?  We claim no novelty for Landau physics in photonics or for optical activity in twisted structures; our results concern the derivation for this particular geometry and its consequences.

One constraint is essential for electromagnetism.  In a Riemann--Cartan description the gauge-invariant electromagnetic field strength remains the exterior derivative~\cite{HehlRMP1976,Shapiro2002,HehlObukhov}
\begin{equation}
F=dA,\qquad F_{\mu\nu}=\partial_\mu A_\nu-\partial_\nu A_\mu,
\label{eq:F=dA}
\end{equation}
so the homogeneous Maxwell equation is the identity $dF=0$.  Replacing the exterior derivative by a torsionful covariant derivative changes the equations, because torsion then appears in the antisymmetrized derivative and in commutators of covariant derivatives~\cite{Puntigam1997,HehlObukhov}.  In minimal Maxwell theory, the effective metric enters through the Hodge operator and the associated Levi--Civita derivative; there is no direct coupling to a separately chosen torsionful connection.  In the defect geometry used below, the profile $f(\rho)$ that enters the metric also has the standard Riemann--Cartan interpretation of a screw-dislocation distribution, but this interpretation plays no role in the electromagnetic dynamics.  A direct torsion coupling would require an additional constitutive or nonminimal term; in fact, a torsion coupling inserted into Maxwell's equations is equivalent to an inhomogeneous isotropic chiral medium with Tellegen and Pasteur parameters~\cite{Horsley2011}.  We introduce no such term.

An earlier analysis by the present author attributed a radius-dependent circular splitting and radial Stokes textures to a uniform screw-dislocation density~\cite{SilvaPRA2026}.  A later reexamination found several independent problems in that reduction.  Although the field strength was written as $F=dA$, the homogeneous equation was recast with a torsionful antisymmetrized covariant derivative, which is not equivalent to the exterior identity $dF=0$ when torsion is nonzero [Eq.~\eqref{eq:homdifference}], and the inhomogeneous equation was written with the torsionful divergence $\widetilde\nabla_\mu F^{\mu\nu}$, which differs from the metric divergence by the torsion terms of Eq.~\eqref{eq:divdifference}.  Within the same approximation, contortion contributions were retained after Levi--Civita terms of the same order had been discarded, even though the corresponding axial transverse blocks cancel for the uniform defect connection used there.  In addition, the coordinate component $K_{\rho\phi z}=\Omega_g\rho$ was interpreted too directly as a local physical coupling, although the orthonormal component is $K_{\hat\rho\hat\phi\hat z}=\Omega_g$, and the propagation scale $\Omega_o\equiv k_0\Omega_g$ was introduced rather than derived from the minimally coupled metric-Maxwell equations.  Together, these points invalidate the radius-dependent local optical splitting on which the structured-polarization claim was based.  Other applications of the same geometry share some of these assumptions or have different, narrower problems; they are discussed in Sec.~\ref{sec:relation_prior}.

Here we do not correct that calculation term by term.  Instead, we rederive the metric-Maxwell problem and the beam equations from the beginning and ask what the screw geometry does to electromagnetic propagation.  Our main results are as follows.  First, geometric optics gives a common principal light cone for both polarizations and a uniform parallel-transport rotation for axial rays.  Second, the full vector Maxwell system leads in the paraxial regime to a two-component Hamiltonian consisting of a Landau-type orbital operator plus a helicity term.  For a uniform defect density, the evolution operator is a tensor product of an orbital propagator and a global polarization rotation.  We derive the closed Landau kernel for an arbitrary input and prove that, at this order, spatial--polarization separability is preserved.  The same spectrum predicts a rigid rotation of balanced $\pm\ell$ OAM petal modes at the geometric rate $-\Omega_g$, exactly matching the rotation rate of a uniform linear polarization.  Third, for an arbitrary smooth axisymmetric density $\sigma(\rho)$, we derive first the metric function and then the paraxial equations.  We show that a uniform density is the only case in which the leading bulk propagator factorizes into orbital and polarization parts for arbitrary inputs.  For a graded density, the beam-weighted characteristic function $\langle e^{-ibz\sigma}\rangle$ controls the local-phase polarization coherence: its first cumulant gives the mean rotation, its variance gives the universal initial texture slope, and higher cumulants organize subsequent local-phase corrections.  Diffraction generates local $S_3$ only at second order, while the integrated circular Stokes component remains exactly conserved in the lossless leading-paraxial theory.  Numerical propagation for a smooth saturating density and for three further bounded profiles confirms these laws.  Writing the transverse Hamiltonian in Pauli form gives its modes with zero propagation shift, and the reduced polarization coherence matrix relates spatial--polarization nonseparability to the measured Stokes textures.  Exact Maxwell eigenmodes for a graded profile test the accuracy of the approximation.  In the leading-order paraxial theory, a uniform defect density thus produces only geometric polarization transport and orbital dynamics, whereas density gradients produce polarization textures across the beam.

Three of these results go beyond known Landau physics and polarization rotation: a factorization theorem that excludes polarization texture for any input in a uniform density; a relation between the density distribution sampled by the beam and the local-phase coherence, which yields an initial texture slope independent of the profile; and a set of vector-level checks (Pauli form, normalizable zero-shift modes, exact Maxwell modes) that bound the approximation.  They also suggest a differential experiment.  A uniform polarization rotation and a spatially varying Stokes field are measured in different ways: the first depends on how the optical frames are registered at the faces of a finite sample, whereas the second can be obtained from calibrated full-Stokes images taken with linear and circular analyzers, for example with digital or polarization-camera methods~\cite{Singh2020,CoxRosales2023}.  Two profiles with the same beam-weighted mean twist can then be told apart.  At leading paraxial order, the uniform one produces no Stokes texture for a separable input, whereas the graded one produces texture with a universal nonzero initial slope whenever the beam samples a nonzero density variance, together with local circular polarization at fixed integrated $S_3$.  Section~\ref{sec:experimental_protocol} develops these points into a measurement scheme, a criterion for choosing the beam waist, and a numerical example.

The scope is distinct from the source-driven model of Ref.~\cite{SilvaAS13014}, where the same screw geometry is supplemented by an independent Carroll--Field--Jackiw-type material response coefficient.  Here no such constitutive axial coupling is introduced.  This lets us isolate the optical effects of the geometry itself and identify which observables require spatial inhomogeneity or additional material physics.

\section{Screw-dislocation geometry and Maxwell equations}
\label{sec:geometry}

\subsection{Assumptions}

The propagation problem has two dynamical inputs: the effective metric $g_{\mu\nu}[f]$ and the minimally coupled Maxwell action.  A chosen dislocation-density profile fixes $f(\rho)$ through the continuum defect model, while the incident beam and the definitions of the measured quantities enter as boundary conditions.  We add no helicity term, gauge potential, Landau coupling, or polarization-texture law by hand.  The Riemann--Cartan connection introduced below serves only to identify the torsion and contortion associated with the defect geometry; it does not enter the Maxwell equations.  All propagation equations follow from the metric Maxwell system, and each approximation is stated together with the small parameter that controls it.

\subsection{Metric, Burgers flux, and defect density}

We use signature $(-,+,+,+)$ and $x^0=ct$.  The fundamental geometric input is the axisymmetric effective metric
\begin{equation}
 ds^2=-(dx^0)^2+d\rho^2+\rho^2d\phi^2+[dz+f(\rho)d\phi]^2.
\label{eq:metric}
\end{equation}
With $f\propto\rho^2$, Eq.~\eqref{eq:metric} is the effective metric of a uniform distribution of parallel screw dislocations~\cite{SilvaNetto2008,BakkeMoraes2012} in the geometric theory of defects~\cite{Katanaev1992,Katanaev2005,Kleinert1989}.  The same metric underlies earlier work by the author and collaborators on electronic and optical problems in this geometry~\cite{SilvaAnnalenQD2026,PereiraSilva2026,SilvaOQE2026,SilvaPRA2026,SilvaAS13014}.  A convenient orthonormal coframe realizing Eq.~\eqref{eq:metric} is
\begin{align}
\vartheta^{\hat0}&=dx^0, &
\vartheta^{\hat1}&=d\rho,\\
\vartheta^{\hat2}&=\rho\,d\phi, &
\vartheta^{\hat3}&=dz+f(\rho)d\phi.
\label{eq:coframe}
\end{align}
The inverse metric and the determinant are
\begin{align}
g^{\phi\phi}&=\rho^{-2}, &
g^{\phi z}&=-\frac{f}{\rho^2},\nonumber\\
g^{zz}&=1+\frac{f^2}{\rho^2}, &
\sqrt{-g}&=\rho,
\label{eq:inverse_det}
\end{align}
with $g^{00}=-1$ and $g^{\rho\rho}=1$.  We call $f(\rho)$ the geometric profile.  Its relation to the screw-dislocation density is best obtained from the Burgers flux~\cite{Katanaev2005,Kleinert1989}, not from a coordinate component of the metric.  We take $b>0$ to be the magnitude of the Burgers vector carried by one elementary screw dislocation, while $\sigma(\rho)$ is a signed areal number density: its sign records the two possible Burgers orientations relative to the chosen $z$ axis.  The net signed number of dislocations crossing a transverse disk $D_\rho$ is
\begin{equation}
N(\rho)=2\pi\int_0^\rho \sigma(\rho')\rho'\,d\rho'.
\label{eq:N_sigma}
\end{equation}
In general, a singular screw line on the symmetry axis may add a constant Burgers flux.  Writing $f_0=f(0)$, a loop at fixed $z$ obeys
\begin{equation}
2\pi\,[f(\rho)-f_0]=bN(\rho).
\label{eq:Burgers_flux}
\end{equation}
We consider the regular case without a singular axial core, so that $f_0=0$.  Stokes' theorem then gives the defect map
\begin{equation}
f(\rho)=b\int_0^\rho\sigma(\rho')\rho'\,d\rho',\qquad
\sigma(\rho)=\frac{f'(\rho)}{b\rho}.
\label{eq:defect_map}
\end{equation}
A nonzero $f_0=B_{\rm axis}/(2\pi)$ can be restored when an isolated axial screw of signed Burgers flux $B_{\rm axis}$ is present; it does not alter the local continuum relation $f'/\rho=b\sigma$ away from the axis.  This relation belongs to the defect model: it specifies the profile that generates the metric and is not an electromagnetic constitutive relation.  We define the local geometric twist
\begin{equation}
\Omega(\rho)\equiv\frac{f'(\rho)}{2\rho}=\frac{b\sigma(\rho)}{2}.
\label{eq:Omega}
\end{equation}
For a homogeneous defect density, the case studied in Refs.~\cite{SilvaNetto2008,BakkeMoraes2012},
\begin{align}
\sigma(\rho)&=\sigma_0, & \Omega(\rho)&=\Omega_g,\nonumber\\
f(\rho)&=\Omega_g\rho^2, & \Omega_g&=\frac{b\sigma_0}{2}.
\label{eq:uniform}
\end{align}
The factor $\rho$ in Eq.~\eqref{eq:defect_map} is the physical cylindrical area factor: $dA=\rho\,d\rho\,d\phi$.  It does not represent a radial variation of the local density.

The metric fixes its Levi--Civita connection uniquely but does not, by itself, fix a torsionful Riemann--Cartan connection.  To connect the metric profile with the standard continuum-defect interpretation~\cite{Bilby1955,Kroner1981,Katanaev1992,SilvaNetto2008}, we adopt the metric-compatible flat defect connection with $\widetilde\omega^{\hat1}{}_{\hat2}=-d\phi$ as its only independent nonzero component.  We use this connection only in the present subsection to identify torsion and contortion.  With the convention $T^{\hat a}=\tfrac12T^{\hat a}{}_{\hat b\hat c}\vartheta^{\hat b}\wedge\vartheta^{\hat c}$, Cartan's first structure equation gives
\begin{align}
T^{\hat3}
&=\frac{f'(\rho)}{\rho}\,\vartheta^{\hat1}\wedge\vartheta^{\hat2}
=b\sigma(\rho)\,\vartheta^{\hat1}\wedge\vartheta^{\hat2}\nonumber\\
&=b\sigma(\rho)\rho\,d\rho\wedge d\phi.
\label{eq:torsion}
\end{align}
Hence the orthonormal coefficient $f'/\rho=b\sigma=2\Omega$ is the local physical defect strength, whereas the coordinate coefficient $T^z{}_{\rho\phi}=f'=b\sigma\rho$ contains the scale factor associated with the angular coordinate.  For the homogeneous distribution, the former is constant while the latter grows linearly with $\rho$ only because $\vartheta^{\hat2}=\rho d\phi$.  The medium is radially inhomogeneous only when $f'(\rho)/\rho$, or equivalently $\sigma(\rho)$, varies with $\rho$.

If the axis belongs to the material domain and $\sigma(\rho)$ is the restriction of a smooth Cartesian scalar $\sigma(\sqrt{x^2+y^2})$, regularity further requires an even near-axis expansion,
\begin{align}
\sigma(\rho)&=\sigma_0+\sigma_2\rho^2+O(\rho^4),\nonumber\\
f(\rho)&=\frac{b\sigma_0}{2}\rho^2+\frac{b\sigma_2}{4}\rho^4+O(\rho^6).
\label{eq:regular_axis_expansion}
\end{align}
The leading smooth radial variation of the density near the axis is quadratic.  A term proportional to $\rho$ can be used on an annular domain that excludes the origin, but it is not differentiable as a rotationally symmetric Cartesian scalar at $\rho=0$.

Solving the torsion-free Cartan equation $d\vartheta^{\hat a}+\omega^{\hat a}{}_{\hat b,\rm LC}\wedge\vartheta^{\hat b}=0$ for the coframe in Eq.~\eqref{eq:coframe} gives
\begin{align}
\omega^{\hat1}{}_{\hat2,\rm LC}&=-\frac{1}{\rho}\vartheta^{\hat2}-\Omega(\rho)\vartheta^{\hat3},\\
\omega^{\hat1}{}_{\hat3,\rm LC}&=-\Omega(\rho)\vartheta^{\hat2},\\
\omega^{\hat2}{}_{\hat3,\rm LC}&=+\Omega(\rho)\vartheta^{\hat1},
\end{align}
and the contortion one-forms~\cite{HehlRMP1976} are
\begin{align}
K^{\hat1}{}_{\hat2}&=+\Omega(\rho)\vartheta^{\hat3},\\
K^{\hat1}{}_{\hat3}&=+\Omega(\rho)\vartheta^{\hat2},\\
K^{\hat2}{}_{\hat3}&=-\Omega(\rho)\vartheta^{\hat1}.
\end{align}
In particular,
\begin{equation}
K_{\hat1\hat2\hat3}=\Omega_g,\qquad
K_{\rho\phi z}=\Omega_g\rho
\end{equation}
for a uniform defect density.  Along the axial frame direction, the transverse Levi--Civita and contortion blocks are equal and opposite.  One therefore cannot keep the contortion term while discarding the Levi--Civita contribution of the same order.

The uniform geometry is homogeneous; its spatial part is the left-invariant metric of the three-dimensional Heisenberg group, the model geometry Nil of Ref.~\cite{Scott1983}, with constant spatial scalar curvature $R^{(3)}=-2\Omega_g^2$.  In Cartesian coordinates,
\begin{equation}
\vartheta^{\hat3}=dz+\Omega_g(x\,dy-y\,dx),
\end{equation}
so $T^{\hat3}=2\Omega_g\,dx\wedge dy$ is constant.  A local property of this homogeneous geometry cannot depend on the distance to an arbitrarily chosen axis.  Beam and mode profiles can depend on $\rho$, because the state and its boundary conditions select an axis.

\subsection{Gauge-invariant Maxwell equations}

We adopt the minimal action~\cite{LandauLifshitzFields,MTW1973,HehlObukhov}
\begin{equation}
S[A]=\int d^4x\sqrt{-g}\left(-\frac14 F_{\mu\nu}F^{\mu\nu}-A_\mu j^\mu\right),
\qquad F=dA.
\label{eq:action}
\end{equation}
Using $\delta F_{\mu\nu}=2\nabla_{[\mu}\delta A_{\nu]}$ and integrating by parts, variation gives
\begin{equation}
\nabla_\mu F^{\mu\nu}=j^\nu,\qquad dF=0,
\label{eq:maxwell_cov}
\end{equation}
where $\nabla$ is the Levi--Civita derivative of the metric.  Current conservation follows both from gauge invariance of the source term and from the antisymmetry of $F$.

The torsionful divergence differs from the metric one~\cite{HehlRMP1976,Shapiro2002}.  For the geometry above,
\begin{equation}
\widetilde\nabla_\mu F^{\mu\nu}-\nabla_\mu F^{\mu\nu}
=\delta^\nu_z f'(\rho)F^{\rho\phi},
\label{eq:divdifference}
\end{equation}
for the stated connection-index convention.  A similar identity holds for the homogeneous equation.  For any antisymmetric $F_{\mu\nu}$, with the same convention,
\begin{equation}
\widetilde\nabla_{[\lambda}F_{\mu\nu]}
=\partial_{[\lambda}F_{\mu\nu]}-T^\sigma{}_{[\lambda\mu}F_{|\sigma|\nu]}.
\label{eq:homdifference}
\end{equation}
For the geometry above, the torsion term has the single independent component $T^\sigma{}_{[0\rho}F_{|\sigma|\phi]}=\tfrac13f'(\rho)F_{z0}$.  Imposing $\widetilde\nabla_{[\lambda}F_{\mu\nu]}=0$ in place of $dF=0$ thus adds a term proportional to $f'(\rho)F_{z0}$ to the axial component of Faraday's law and leaves the other components unchanged.  For this reason we work with Eq.~\eqref{eq:maxwell_cov}.

With the curvature convention $[\nabla_\mu,\nabla_\nu]A^\alpha=R^\alpha{}_{\beta\mu\nu}A^\beta$, expanding $\nabla_\mu F^{\mu\nu}=0$ and imposing the Levi--Civita Lorenz gauge $\nabla_\mu A^\mu=0$ gives~\cite{MTW1973}
\begin{equation}
\nabla_\alpha\nabla^\alpha A^\nu-R^\nu{}_{\mu}A^\mu=0.
\label{eq:wave}
\end{equation}

\subsection{Equivalent medium and scope of the model}
\label{sec:metric_medium}

Following Plebanski, the metric Maxwell system can also be read as Maxwell theory in flat laboratory space filled with an effective transformation-optics medium~\cite{Plebanski1960,Leonhardt2006,Pendry2006}.  For a static metric with $g_{0i}=0$ and $g_{00}=-1$, the constitutive relations contain no magnetoelectric vector and are
\begin{equation}
D^i=\epsilon_0\,\varepsilon^{ij}E_j,\qquad
B^i=\mu_0\,\mu^{ij}H_j,
\label{eq:plebanski_constitutive}
\end{equation}
with
\begin{equation}
\varepsilon^{ij}=\mu^{ij}=\sqrt{-g}\,g^{ij},\qquad \gamma^{ij}=0.
\label{eq:plebanski_metric}
\end{equation}
Here $\varepsilon^{ij}$ and $\mu^{ij}$ are the usual contravariant constitutive tensor densities associated with the chosen spatial coordinates.  For Eq.~\eqref{eq:metric},
\begin{equation}
\varepsilon^{ij}=\mu^{ij}=
\begin{pmatrix}
\rho&0&0\\
0&\rho^{-1}&-f/\rho\\
0&-f/\rho&\rho+f^2/\rho
\end{pmatrix}_{(\rho,\phi,z)}.
\label{eq:metric_medium_matrix}
\end{equation}
The factors of $\rho$ in Eq.~\eqref{eq:metric_medium_matrix} do not indicate a local inhomogeneity: vacuum written in cylindrical coordinates has the same coordinate-density factors.  The geometry enters through the terms containing $f(\rho)$.  These terms still depend on $\rho$ in an orthonormal cylindrical laboratory frame; for uniform density
\begin{equation}
\varepsilon^{\hat\imath\hat\jmath}=\mu^{\hat\imath\hat\jmath}=
\begin{pmatrix}
1&0&0\\
0&1&-\Omega_g\rho\\
0&-\Omega_g\rho&1+\Omega_g^2\rho^2
\end{pmatrix}_{(\hat\rho,\hat\phi,\hat z)},
\label{eq:metric_medium_lab}
\end{equation}
so the principal axes of the analogue medium tilt more and more with distance from the chosen axis.  This dependence is the transformation-optics counterpart of the geometric gauge potential $\Ag=(f/\rho)\hat{\bm\phi}$ introduced in Sec.~\ref{sec:paraxial}, and it is removed by the helical isometries of Appendix~\ref{app:geometry}, which combine transverse translations with longitudinal shifts.  Local material quantities must therefore be defined through combinations invariant under these longitudinal shifts, such as $\nabla_\perp\times\Ag=f'/\rho=2\Omega$, or through components in the adapted coframe $\vartheta^{\hat a}$, in which $\varepsilon=\mu=\I$ and the geometry appears only in $d\vartheta^{\hat3}=2\Omega\,\vartheta^{\hat1}\wedge\vartheta^{\hat2}$.

Equations~\eqref{eq:plebanski_constitutive}--\eqref{eq:metric_medium_matrix} also show the limits of the model.  The geometry alone gives an impedance-matched metric medium with no independent Pasteur, Tellegen, or Carroll--Field--Jackiw coefficient.  A nonmagnetic natural crystal, with $\mu^{ij}=\delta^{ij}$, does not satisfy $\varepsilon^{ij}=\mu^{ij}$ even approximately, and a real dislocated crystal may also show photoelasticity, intrinsic birefringence, absorption, spatial dispersion, or other microscopic responses that Eq.~\eqref{eq:metric} does not determine.  Predictions for a specific material require these effects to be modeled separately.  The results below apply to the metric Maxwell model itself, or to an analogue medium designed to realize its constitutive tensors, and give the part of the optical response that comes from the geometry alone.

\section{Maxwell equations in the orthonormal frame}
\label{sec:vector}

The dual frame is
\begin{equation}
e_{\hat1}=\partial_\rho,\qquad
e_{\hat2}=\rho^{-1}\partial_\phi-\frac{f(\rho)}{\rho}\partial_z,
\qquad e_{\hat3}=\partial_z.
\end{equation}
For harmonic dependence $e^{i(m\phi+k_z z-\omega t)}$, define
\begin{equation}
q(\rho)=\frac{m-f(\rho)k_z}{\rho}.
\label{eq:q}
\end{equation}
For the physical one-form $U^\flat=U_\rho\vartheta^{\hat1}+U_\phi\vartheta^{\hat2}+U_z\vartheta^{\hat3}$, expanding $dU^\flat$ in the orthonormal two-form basis gives the curl
\begin{align}
(\Cf\mathbf U)_\rho&=iqU_z-ik_zU_\phi,\\
(\Cf\mathbf U)_\phi&=ik_zU_\rho-U_z',\\
(\Cf\mathbf U)_z&=U_\phi'+\frac{U_\phi}{\rho}-iqU_\rho+\frac{f'}{\rho}U_z,
\label{eq:curl}
\end{align}
and the divergence is
\begin{equation}
\Df\mathbf U=U_\rho'+\frac{U_\rho}{\rho}+iqU_\phi+ik_zU_z.
\label{eq:div}
\end{equation}
By construction, $\Df(\Cf\mathbf U)=0$.  The term $(f'/\rho)U_z=2\Omega U_z$ in Eq.~\eqref{eq:curl} couples different field components and has no counterpart in a componentwise scalar Laplace--Beltrami treatment.

With $k_0=\omega/c$ and $\mathbf B=c\mathbf B_{\rm SI}$, the source-free system is
\begin{equation}
\Cf\mathbf E=ik_0\mathbf B,\qquad
\Cf\mathbf B=-ik_0\mathbf E,
\label{eq:maxwell_harmonic}
\end{equation}
with the two Gauss constraints $\Df\mathbf E=\Df\mathbf B=0$.

\section{Geometric optics}
\label{sec:wkb}

Let
\begin{equation}
A^\mu=(a^\mu+\epsilon b^\mu+\cdots)e^{iS/\epsilon},\qquad k_\mu=\partial_\mu S.
\end{equation}
Substitution into Eq.~\eqref{eq:wave} gives, at leading order~\cite{MTW1973},
\begin{equation}
k_\mu k^\mu=0,\qquad k_\mu a^\mu=0.
\label{eq:eikonal}
\end{equation}
Both polarizations therefore propagate on the same null cone.  At the next order,
\begin{equation}
2k^\alpha\nabla_\alpha a^\mu+(\nabla_\alpha k^\alpha)a^\mu=0.
\end{equation}
Writing $a^\mu=\mathcal A f^\mu$ and absorbing the expansion of the ray congruence into $\mathcal A$ gives
\begin{equation}
k^\alpha\nabla_\alpha f^\mu=0.
\label{eq:transport}
\end{equation}

For an axial ray in the uniform geometry, the transverse components satisfy
\begin{equation}
\frac{d}{dz}
\begin{pmatrix}f_{\hat1}\\f_{\hat2}\end{pmatrix}
=\Omega_g
\begin{pmatrix}0&1\\-1&0\end{pmatrix}
\begin{pmatrix}f_{\hat1}\\f_{\hat2}\end{pmatrix}.
\end{equation}
Hence, for a polarization initially along $\hat{\mathbf e}_1$,
\begin{equation}
f_{\hat1}=\cos(\Omega_g z),\qquad
f_{\hat2}=-\sin(\Omega_g z),
\end{equation}
and the signed rotation relative to the transverse lattice frame is
\begin{equation}
\theta_{\rm geom}(L)=-\Omega_g L.
\label{eq:rotation}
\end{equation}
The sign depends on the orientation and circular-basis conventions; the magnitude $|\Omega_g|L$ does not.  Equation~\eqref{eq:rotation} gives the transport angle relative to the geometry-adapted transverse frame; it is not by itself what a pair of laboratory polarizers outside a finite sample would read.  If the entrance and exit analyzer frames differ from the adapted frame by angles $\delta\theta_{\rm in}$ and $\delta\theta_{\rm out}$, the measured angle also contains these offsets,
\begin{equation}
\theta_{\rm meas}=-\Omega_gL+\delta\theta_{\rm out}-\delta\theta_{\rm in}.
\label{eq:measured_rotation}
\end{equation}
The bulk transport angle can be identified with a laboratory analyzer rotation only once the matching at the interfaces is specified.  Equation~\eqref{eq:measured_rotation} assumes polarization-preserving, impedance-matched interfaces whose only effect is the registration of the transverse frames.  For a general interface, the entrance and exit Jones matrices must be included, and no simple additive formula applies.  For a single axial ray, denote by $J_{\rm in}$ the laboratory-to-adapted-frame matching matrix and by $J_{\rm out}$ the adapted-to-laboratory matching matrix.  A laboratory analyzer represented by a unit Jones vector $\mathbf a$ then measures, up to the overall transmitted intensity,
\begin{equation}
I_{\mathbf a}\ \propto\
\left|\mathbf a^\dagger J_{\rm out}
 e^{i\Omega_gL\sigma_3}J_{\rm in}\chi_{\rm lab,in}\right|^2.
\label{eq:interface_jones}
\end{equation}
This requires the circular basis and the transverse registration at each surface.  Fresnel reflection, polarization conversion, absorption, or spatially varying interfaces require their own Jones operators (and, for beams, a treatment of the orbital modes).  A frame-transport angle alone is not a prediction for an external polarimeter.

In the circular basis
\begin{equation}
\mathbf e_\pm=\frac{\mathbf e_x\pm i\mathbf e_y}{\sqrt2},\qquad
\psi_\pm=\frac{E_x\mp iE_y}{\sqrt2},
\label{eq:circular}
\end{equation}
Eq.~\eqref{eq:rotation} corresponds to $\psi_\pm\propto e^{\pm i\Omega_g L}$ relative to the carrier, i.e., to a transport-phase difference of magnitude $2\Omega_g L$.  This is a geometric phase beyond eikonal order; the light cone itself is not split.  It is the finite-sample form of the general observer and frame dependence of polarization transport~\cite{Santana2020}.

\subsection{Polarization conventions and Stokes parameters}
\label{sec:stokes_conventions}

Because the sign of $S_3$ depends on the time-harmonic and circular-basis conventions, we derive the Stokes formulas used below from the analyzer intensities.  With the $e^{-i\omega t}$ convention, write the coherent transverse analytic field as
\begin{equation}
\mathbf E_\perp=E_x\mathbf e_x+E_y\mathbf e_y,
\qquad
\mathbf J=\begin{pmatrix}E_x\\E_y\end{pmatrix},
\label{eq:jones_linear}
\end{equation}
where $\mathbf J$ is the Jones vector.  In common intensity units, ideal analyzers along $x$, $y$, $+45^\circ$, $-45^\circ$, and the two circular basis states measure
\begin{align}
I_x&=|E_x|^2, & I_y&=|E_y|^2,\nonumber\\
I_{45}&=\frac{|E_x+E_y|^2}{2}, &
I_{135}&=\frac{|E_x-E_y|^2}{2},\nonumber\\
I_+&=|\psi_+|^2, & I_-&=|\psi_-|^2.
\label{eq:analyzer_intensities}
\end{align}
We define the Stokes fields by the analyzer differences~\cite{SalehTeich2019},
\begin{align}
S_0&=I_x+I_y, & S_1&=I_x-I_y,\nonumber\\
S_2&=I_{45}-I_{135}, & S_3&=I_+-I_-.
\label{eq:stokes_operational}
\end{align}
Substituting Eqs.~\eqref{eq:circular} and \eqref{eq:analyzer_intensities} gives
\begin{align}
S_0&=|E_x|^2+|E_y|^2,\nonumber\\
S_1&=|E_x|^2-|E_y|^2,\nonumber\\
S_2&=2\operatorname{Re}(E_xE_y^*),\nonumber\\
S_3&=-2\operatorname{Im}(E_xE_y^*).
\label{eq:stokes_linear}
\end{align}
Equivalently, the coherence matrix of the deterministic Jones field is
\begin{equation}
\mathbf J\mathbf J^\dagger=
\frac12\begin{pmatrix}
S_0+S_1 & S_2-iS_3\\
S_2+iS_3 & S_0-S_1
\end{pmatrix}.
\label{eq:coherency_linear}
\end{equation}
Changing to the circular amplitudes in Eq.~\eqref{eq:circular} yields
\begin{align}
S_0&=|\psi_+|^2+|\psi_-|^2,\nonumber\\
S_3&=|\psi_+|^2-|\psi_-|^2,\nonumber\\
S_1+iS_2&=2\psi_+^*\psi_-.
\label{eq:stokes_circular}
\end{align}
These equations fix the helicity sign used throughout.  For the fully coherent fields considered here, they also imply
\begin{equation}
S_0^2=S_1^2+S_2^2+S_3^2.
\label{eq:stokes_pure_identity}
\end{equation}
Finally, a linearly polarized field at angle $\theta$ has $\mathbf J=u(\cos\theta,\sin\theta)^T$, so
\begin{equation}
S_1=S_0\cos2\theta,\qquad
S_2=S_0\sin2\theta,\qquad S_3=0,
\end{equation}
and hence its local orientation is
\begin{equation}
\psi=\frac12\arg(S_1+iS_2)\quad (\mathrm{mod}\;\pi).
\label{eq:stokes_angle}
\end{equation}
The polarization angles and circular-power differences used below follow from the analyzer projections and the circular basis of Eq.~\eqref{eq:circular}.

\section{Paraxial propagation in a uniform defect density}
\label{sec:paraxial}

\subsection{Paraxial equation}

Starting from $\Cf^2\mathbf E=k_0^2\mathbf E$, projecting onto the transverse orthonormal components, and using Gauss's law to eliminate $E_z$ gives the leading paraxial equation (Appendix~\ref{app:paraxial})
\begin{equation}
 i\partial_z\Psi=
 \left[-\frac{1}{2k_0}(\nabla_\perp-i k_0\Ag)^2\I
 -\Omega(\rho)\sigma_3\right]\Psi,
\label{eq:paraxial}
\end{equation}
where
\begin{align}
\Psi&=(\psi_+,\psi_-)^T,\nonumber\\
\Ag&=\frac{f(\rho)}{\rho}\hat{\bm\phi},\qquad
\nabla_\perp\times\Ag=2\Omega(\rho)\hat{\mathbf z}.
\label{eq:Ag}
\end{align}
The paraxial ordering can be stated in a coordinate-invariant way with
\begin{equation}
D_\perp=\nabla_\perp-i k_0\Ag,
\qquad
\epsilon_{\rm cov}\equiv
\frac{\|D_\perp\Psi\|}{k_0\|\Psi\|}\ll1,
\label{eq:covariant_paraxial_ordering}
\end{equation}
together with $|\Omega|/k_0\ll1$ and transverse variation on scales of order the beam width $w$ or longer.  Under the change of longitudinal coordinate $z'=z+\chi(x,y)$ one has $\Ag'=\Ag-\nabla_\perp\chi$ and, in a fixed forward Fourier sector, $\Psi'=e^{-ik_0\chi}\Psi$, so $D_\perp'\Psi'=e^{-ik_0\chi}D_\perp\Psi$.  We call this freedom in the choice of the longitudinal coordinate the coordinate gauge of $\Ag$.  It is distinct from the electromagnetic gauge transformation $A_\mu\to A_\mu+\partial_\mu\Lambda$, which leaves $F$ unchanged.  In the regular axisymmetric coordinate gauge of Eq.~\eqref{eq:Ag}, the usual estimates $(k_0w)^{-1}\ll1$ and $|\Ag|\ll1$ are sufficient, but the second depends on the coordinate gauge; Eq.~\eqref{eq:covariant_paraxial_ordering} is the invariant condition.  Equation~\eqref{eq:paraxial} is the leading-order propagation equation and does not replace the full vector Maxwell system.

For a uniform defect density, $f=\Omega_g\rho^2$, so
\begin{equation}
H_p=\left[-\frac{\nabla_\perp^2}{2k_0}-\Omega_gL_z+
\frac{k_0\Omega_g^2\rho^2}{2}\right]\I-\Omega_g\sigma_3,
\label{eq:Huniform}
\end{equation}
where $L_z=-i\partial_\phi$.  The orbital part has the form of the two-dimensional Landau Hamiltonian in symmetric gauge~\cite{LandauLifshitzQM}, with effective field $\nabla_\perp\times\Ag=2\Omega_g\hat{\mathbf z}$.  The $\Ag^2$ term must be kept on the confinement scale set by the geometry.

\subsection{Landau propagator}

For a uniform defect density, the orbital Hamiltonian in Eq.~\eqref{eq:Huniform} can be written as
\begin{equation}
H_{\rm orb}=H_{\rm osc}-\Omega_g L_z,
\qquad
H_{\rm osc}=\frac{\mathbf p_\perp^2}{2k_0}+\frac{k_0\Omega_g^2\rho^2}{2}.
\label{eq:Hosc}
\end{equation}
Because $[H_{\rm osc},L_z]=0$, its propagator follows from the two-dimensional Mehler kernel~\cite{Kleinert2009} followed by a rigid rotation.  Away from the caustic planes $\sin(\Omega_g z)=0$, the result is
\begin{align}
\mathcal K_{\Omega}(\mathbf r,\mathbf r';z)
={}&\frac{k_0\Omega_g}{2\pi i\sin(\Omega_g z)}
\exp\Bigg[\frac{i k_0\Omega_g}{2}\cot(\Omega_g z)
|\mathbf r-\mathbf r'|^2\nonumber\\
&\hspace{25mm}-i k_0\Omega_g(\mathbf r\times\mathbf r')_z\Bigg].
\label{eq:Landau_kernel}
\end{align}
The free-space Fresnel kernel is recovered continuously as $\Omega_g\rightarrow0$.  Direct differentiation also gives
\begin{equation}
i\partial_z\mathcal K_{\Omega}=H_{\rm orb}\mathcal K_{\Omega},
\end{equation}
while the factorization in Eq.~\eqref{eq:Hosc} guarantees the unitary composition law.  Any scalar transverse input evolves as
\begin{equation}
u_z(\mathbf r)=\int d^2r'\,\mathcal K_{\Omega}(\mathbf r,\mathbf r';z)u_0(\mathbf r').
\label{eq:general_prop}
\end{equation}
Equation~\eqref{eq:Landau_kernel} applies to any input beam, not only to the centered Gaussian considered below; it contains diffraction, cyclotron-like rotation, and the confinement produced by the geometry.  The situation is analogous to synthetic magnetic fields for photons~\cite{Schine2016,Guglielmon2021}, except that here the effective field is set by the screw-dislocation geometry instead of a resonator or a strain pattern.  Appendix~\ref{app:propagator} gives the derivation and the short-distance check.

\subsection{Separability for a uniform density}

Write
\begin{equation}
H_p=H_{\rm orb}\I-\Omega_g\sigma_3.
\end{equation}
Because $[H_{\rm orb},\sigma_3]=0$,
\begin{equation}
U(z)=e^{-izH_{\rm orb}}e^{+i\Omega_g z\sigma_3}.
\label{eq:factorization}
\end{equation}
Consider any input whose two circular components share the same spatial profile,
\begin{equation}
\Psi(\mathbf r_\perp,0)=u_0(\mathbf r_\perp)\,\chi_0.
\end{equation}
Equation~\eqref{eq:factorization} gives
\begin{align}
\Psi(\mathbf r_\perp,z)&=u_z(\mathbf r_\perp)\,\chi_z,\nonumber\\
u_z&=e^{-izH_{\rm orb}}u_0,\qquad
\chi_z=e^{+i\Omega_gz\sigma_3}\chi_0.
\label{eq:separable}
\end{align}
The normalized polarization state thus remains independent of $\mathbf r_\perp$.  More generally, Eq.~\eqref{eq:factorization} is a product of unitaries acting on the transverse-mode and polarization spaces, so the Schmidt coefficients of any pure spatial--polarization state are conserved: within the leading paraxial theory, a uniform defect density cannot make a separable input nonseparable.  The factorization holds at leading paraxial order only; the exact Maxwell fields contain longitudinal and radial-scale corrections, which are estimated in Sec.~\ref{sec:maxwell_paraxial_validity}.  For an input with a spatially uniform Jones vector, it follows that no polarization texture is generated.  Using the Stokes convention derived in Eq.~\eqref{eq:stokes_circular}, a linearly polarized input at angle $\theta_0$ gives
\begin{align}
S_0&=|u_z|^2,\\
S_1+iS_2&=S_0e^{2i(\theta_0-\Omega_gz)},\\
S_3&=0.
\label{eq:stokes_uniform}
\end{align}
Thus, within the domain of Eq.~\eqref{eq:paraxial},
\begin{equation}
\psi(\mathbf r_\perp,z)=\theta_0-\Omega_gz
\end{equation}
is spatially uniform.  A uniform defect density can reshape the intensity and the orbital phase through $H_{\rm orb}$, but it cannot turn an initially uniform polarization into radial polarization domains.  This statement concerns the leading-order paraxial bulk theory; interface effects and nonparaxial corrections are separate questions.

\begin{figure}[t]
\includegraphics[width=\columnwidth]{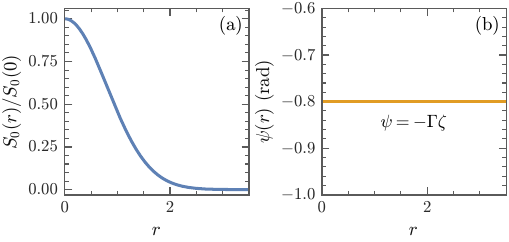}
\caption{Uniform defect density in the leading paraxial theory.  (a) The centered Gaussian intensity $S_0(r)$ varies transversely, but (b) the polarization angle $\psi(r)=-\Gamma\zeta$ is independent of $r$.  The example uses $\Gamma=1$ and $\zeta=0.8$.}
\label{fig:no_texture}
\end{figure}

Figure~\ref{fig:no_texture} illustrates the factorization for a centered Gaussian.  The intensity varies across the beam, while the local polarization angle does not depend on the radius.

\subsection{Rotation of OAM petal patterns}

\begin{figure}[t]
\includegraphics[width=\columnwidth]{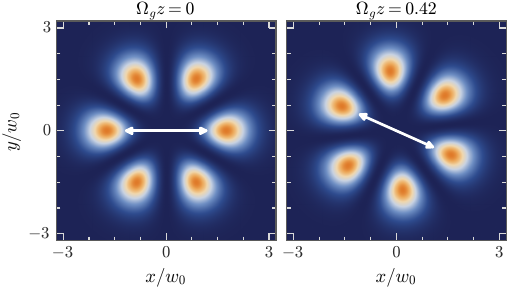}
\caption{Uniform defect density: rigid rotation of a balanced $m=\pm3$ OAM superposition for $\Omega_g>0$.  The petal pattern rotates at $-\Omega_g$, Eq.~\eqref{eq:petal_rate}.  The white double-headed arrows mark the uniform linear-polarization axis at input and output, which rotates at the same geometric rate.  The example uses $\Gamma=1$, with coordinates in units of $w_0$; the $n=0$ radial profile is matched to the Landau scale, so the pattern does not breathe.  Reversing the sign of $\Omega_g$ reverses both rotations.  The co-rotation does not imply spin--orbit conversion because the spatial and polarization propagators remain factorized.}
\label{fig:petals}
\end{figure}

The Landau modes give a second observable that does not involve polarization texture.  Separating $H_{\rm osc}$ in polar coordinates with $u(\rho,\phi)=R(\rho)e^{im\phi}$ and setting $x=k_0|\Omega_g|\rho^2$ reduces the radial equation to the associated-Laguerre equation, as in the Fock--Darwin problem~\cite{Fock1928,Darwin1931}.  Normalizability gives the oscillator eigenvalue $|\Omega_g|(2n+|m|+1)$; subtracting the $\Omega_g L_z$ term gives
\begin{equation}
\beta^{\rm orb}_{nm}=|\Omega_g|(2n+|m|+1)-\Omega_g m.
\label{eq:orb_eigs_general}
\end{equation}
Consider the balanced $n=0$ superposition of counter-rotating OAM modes $m=\pm\ell$.  Their radial profiles are identical because they depend only on $|m|$ and $|\Omega_g|$.  Using Eq.~\eqref{eq:orb_eigs_general}, the two propagation shifts are
\begin{equation}
\beta_{+\ell}=|\Omega_g|(\ell+1)-\Omega_g\ell,
\qquad
\beta_{-\ell}=|\Omega_g|(\ell+1)+\Omega_g\ell.
\end{equation}
Hence, for either sign of $\Omega_g$,
\begin{align}
u_{\ell}(\rho,\phi,z)
&\propto R_{\ell}(\rho)e^{-i|\Omega_g|(\ell+1)z}
\Big[e^{i\ell(\phi+\Omega_g z)}\nonumber\\
&\hspace{28mm}+e^{-i\ell(\phi+\Omega_g z)}\Big]\nonumber\\
&=2R_{\ell}(\rho)e^{-i|\Omega_g|(\ell+1)z}
\cos\!\left[\ell(\phi+\Omega_g z)\right].
\label{eq:petal_rotation}
\end{align}
The $2\ell$-petal intensity pattern therefore rotates rigidly as
\begin{equation}
\phi_{\rm petal}(z)=-\Omega_g z,
\label{eq:petal_rate}
\end{equation}
independently of $\ell$ and with the sense reversed when the screw handedness is reversed.  A uniform linear polarization simultaneously obeys $\theta(z)=\theta_0-\Omega_g z$.  The orbital pattern and the polarization axis rotate together while the state remains separable.  This co-rotation differs from spin--orbit conversion: no weight is transferred between OAM sectors and no spatially varying polarization is produced.  Figure~\ref{fig:petals} illustrates the co-rotation for the balanced $\ell=3$, $n=0$ superposition.

\subsection{Width of a Gaussian beam}

\begin{figure}[t]
\includegraphics[width=\columnwidth]{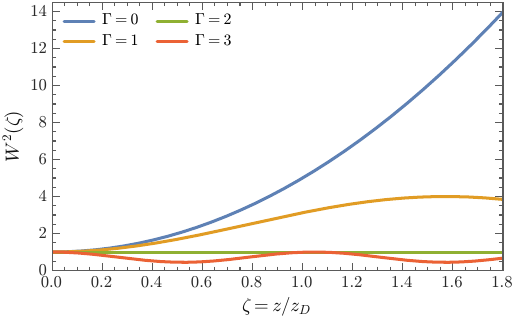}
\caption{Gaussian beam width from Eq.~\eqref{eq:width} for a uniform defect density.  The orbital dynamics are nontrivial even though the normalized polarization stays spatially uniform.  For $|\Gamma|=2$ the width is constant.}
\label{fig:breathing}
\end{figure}

For a centered rotationally symmetric Gaussian, $L_z=0$ and the orbital dynamics reduces to a two-dimensional harmonic oscillator.  Let
\begin{equation}
r=\frac{\rho}{w_0},\qquad
\zeta=\frac{z}{z_D},\qquad
z_D=k_0w_0^2,\qquad
\Gamma=\Omega_g z_D,
\end{equation}
where $z_D\equiv k_0w_0^2$ is the diffraction length used throughout.  For the amplitude convention $u(\rho,0)\propto\exp(-\rho^2/w_0^2)$, the conventional Gaussian Rayleigh range is $z_R^{\rm(std)}=k_0w_0^2/2=z_D/2$; we use the symbol $z_D$ to keep the two lengths apart.  We take $u(r,0)\propto e^{-r^2}$.  For each Cartesian coordinate the oscillator Heisenberg evolution is
\begin{equation}
\hat x(\zeta)=\hat x(0)\cos(\Gamma\zeta)+
\frac{\hat p_x(0)}{\Gamma}\sin(\Gamma\zeta),
\label{eq:heisenberg_width}
\end{equation}
with the continuous $\Gamma\to0$ limit understood.  The input Gaussian has $\langle x^2\rangle_0=1/4$, $\langle p_x^2\rangle_0=1$, and $\langle xp_x+p_xx\rangle_0=0$.  Hence the propagated intensity remains Gaussian,
\begin{equation}
S_0(r,\zeta)=\frac{2}{\pi W^2(\zeta)}
\exp\left[-\frac{2r^2}{W^2(\zeta)}\right],
\end{equation}
with
\begin{equation}
W^2(\zeta)=\cos^2(\Gamma\zeta)+\frac{4}{\Gamma^2}\sin^2(\Gamma\zeta).
\label{eq:width}
\end{equation}
The free-space limit is $W^2\to1+4\zeta^2$ as $\Gamma\to0$.  A simple special case is
\begin{equation}
|\Gamma|=2,\qquad W(\zeta)=1,
\label{eq:stationary}
\end{equation}
for which the Gaussian is matched to the transverse confinement and its width stays constant.  Figure~\ref{fig:breathing} shows $W^2(\zeta)$ for $\Gamma=0$, $1$, $2$, and $3$.

\subsection{Comparison with exact Maxwell modes}

The exact source-free Maxwell equations give an independent check of the spectrum.  From Eq.~\eqref{eq:maxwell_harmonic}, $\Cf^2\mathbf E=k_0^2\mathbf E$, so the source-free solution space decomposes into Beltrami sectors~\cite{Moses1971,SilvaAS13014} $\Cf\mathbf E_\eta=\eta k_0\mathbf E_\eta$, $\eta=\pm1$.  Solving the first two Beltrami components for $E_{\eta,\rho}$ and $E_{\eta,\phi}$ and substituting them into the longitudinal component gives, away from $k_0^2-k_z^2=0$,
\begin{align}
u_\eta''+\frac{1}{\rho}u_\eta'
-\Bigg[&\left(\frac{m}{\rho}-\Omega_gk_z\rho\right)^2
+k_z^2-k_0^2\nonumber\\
&+2\eta\Omega_gk_0\Bigg]u_\eta=0.
\label{eq:radial_exact}
\end{align}
For $\Omega_gk_z>0$, normalizability gives Laguerre--Gauss profiles and
\begin{align}
k_z^2-k_0^2+2\eta\Omega_gk_0+2\Omega_gk_zN_{nm}&=0,\nonumber\\
N_{nm}&=2n+|m|-m+1.
\label{eq:exact_spectrum}
\end{align}
Thus
\begin{equation}
k_{z,\eta}=-\Omega_gN_{nm}+
\sqrt{k_0^2+\Omega_g^2N_{nm}^2-2\eta\Omega_gk_0}.
\end{equation}
Its weak-twist expansion is
\begin{equation}
k_{z,\eta}=k_0-\Omega_g(N_{nm}+\eta)
+\frac{(N_{nm}^2-1)\Omega_g^2}{2k_0}
+O(\Omega_g^3/k_0^2).
\label{eq:exact_expand}
\end{equation}
Comparing with Eq.~\eqref{eq:Huniform} requires one further step.  The longitudinal Laguerre mode carries the index $N_{nm}$, whereas reconstruction of the dominant transverse circular component applies a first-order Landau ladder operator.  For the forward branch with $\Omega_gk_z>0$, the result is
\begin{equation}
N_{\perp}=N_{nm}+2\eta,
\label{eq:Nladder}
\end{equation}
with the same helicity label $\eta$ as the Beltrami sector.  The paraxial transverse propagation shift is then
\begin{equation}
\beta_{\perp}=\Omega_g(N_{\perp}-\eta)
=\Omega_g(N_{nm}+\eta),
\end{equation}
which agrees exactly with the first-order term of Eq.~\eqref{eq:exact_expand}.  No relabeling of helicity is needed; the apparent mismatch is removed by the shift of the transverse Landau index.  For the lowest longitudinal rung with $\eta=-1$, $N_{nm}=1$, Eq.~\eqref{eq:Nladder} would require $N_{\perp}=-1$: the lowering operator annihilates the dominant transverse component, which shows that this longitudinal branch is unphysical.  The physical exceptional family, in contrast, has $E_z=0$ and does not arise from a longitudinal rung; its lowest $\eta=+1$ state has $N_{\perp}=1$ and hence zero paraxial propagation shift.  Appendix~\ref{app:ladder} gives the Laguerre identities, and Appendix~\ref{app:exceptional} discusses the exceptional sector.

\subsection{Accuracy of the paraxial approximation}
\label{sec:maxwell_paraxial_validity}

The exact spectrum also gives an error estimate for the paraxial evolution.  For definiteness take $\Omega_g>0$ and define
\begin{equation}
\gamma=\frac{\Omega_g}{k_0},\qquad
\kappa_{N\eta}=\frac{k_{z,\eta}}{k_0},\qquad
\beta_{\rm ex}=k_0-k_{z,\eta},
\end{equation}
with $N=N_{nm}$.  For the opposite handedness, the same result holds with the azimuthal labels reversed.  Equation~\eqref{eq:exact_spectrum} gives
\begin{equation}
\frac{\beta_{\rm ex}}{k_0}
=1+\gamma N-
\sqrt{1+\gamma^2N^2-2\eta\gamma},
\label{eq:beta_exact_dimless}
\end{equation}
whereas the leading paraxial Hamiltonian gives
\begin{equation}
\frac{\beta_{\rm p}}{k_0}=\gamma(N+\eta).
\label{eq:beta_par_dimless}
\end{equation}
Their difference can be written exactly as
\begin{align}
\frac{\delta\beta}{k_0}
&\equiv\frac{\beta_{\rm p}-\beta_{\rm ex}}{k_0}\nonumber\\
&=\frac{\gamma^2(N^2-1)}
{\sqrt{1+\gamma^2N^2-2\eta\gamma}+1-\eta\gamma}.
\label{eq:delta_beta_exact}
\end{align}
The paraxial propagation shift therefore exceeds the exact one by an amount of order $\gamma^2$ for every nonexceptional mode with $N>1$, and it is exact for $N=1$.  For $N+\eta\neq0$,
\begin{equation}
\frac{\delta\beta}{\beta_{\rm p}}
=\frac{N-\eta}{2}\,\gamma+O(\gamma^2).
\label{eq:relative_beta_error}
\end{equation}
When high Landau rungs are populated, the relevant small parameter is $N\Omega_g/k_0$, not $\Omega_g/k_0$.

For a finite beam, let $Q=k_0w_0$ and $\Gamma=\Omega_g z_D$ with $z_D=k_0w_0^2$.  Since
\begin{equation}
\gamma=\frac{\Gamma}{Q^2},
\end{equation}
the modal phase error accumulated over one diffraction length $z_D$ is
\begin{equation}
\delta\Phi_N(z_D)=\delta\beta\,z_D
=\frac{N^2-1}{2}\frac{\Gamma^2}{Q^2}
+O(Q^{-4}).
\label{eq:phase_error_zD}
\end{equation}
Equation~\eqref{eq:phase_error_zD} gives the admissible range of $\Gamma/Q$ for an input whose highest populated Landau index is known.

A second comparison tests the vector structure of the field, not only the propagation constant.  The exact $\eta=+1$, $n=m=0$ branch has $N=1$ and, for $0<\gamma<1/2$,
\begin{equation}
 k_z=k_0-2\Omega_g.
\end{equation}
Its longitudinal field is $u\propto\exp(-\Omega_g k_z\rho^2/2)$, and reconstructing the field from Maxwell's first-order equations gives the fraction of the norm carried by the longitudinal electric field,
\begin{equation}
f_z=\frac{\|E_z\|^2}{\|\mathbf E\|^2}=2\gamma.
\label{eq:fz_exact}
\end{equation}
The corresponding transverse circular mode has the same $\ell=-1$, $p=0$ Laguerre--Gaussian form (the $N_\perp=3$ Landau state) as the paraxial mode, but with the exact radial scale $\alpha_{\rm ex}=\Omega_gk_z$ instead of $\alpha_{\rm p}=\Omega_gk_0$.  The overlap fidelity of the two normalized transverse profiles is
\begin{equation}
\mathcal F_\perp=
\left[\frac{2\sqrt{\kappa}}{1+\kappa}\right]^4,
\qquad \kappa=1-2\gamma,
\label{eq:profile_fidelity}
\end{equation}
so that $1-\mathcal F_\perp=2\gamma^2+O(\gamma^3)$.  Appendix~\ref{app:maxwell_paraxial_benchmark} derives Eqs.~\eqref{eq:fz_exact} and \eqref{eq:profile_fidelity} from the exact field.

\begin{figure}[t]
\includegraphics[width=\columnwidth]{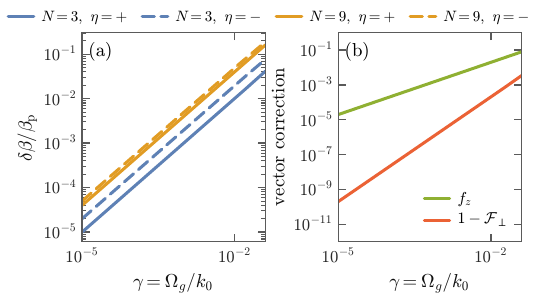}
\caption{Exact uniform Maxwell modes compared with the leading paraxial theory.  (a) Relative error of the paraxial propagation shift for two Landau indices; solid curves, $\eta=+$; dashed curves, $\eta=-$.  The unphysical exceptional branch $N=1$, $\eta=-$ is excluded.  (b) Lowest $\eta=+$, $n=m=0$ mode: longitudinal electric-field fraction $f_z$ and transverse-profile infidelity $1-\mathcal F_\perp$.  All these quantities vanish as $\gamma=\Omega_g/k_0\to0$.}
\label{fig:maxwell_paraxial_validity}
\end{figure}

Figure~\ref{fig:maxwell_paraxial_validity} summarizes both comparisons.  For example, $\Gamma=2$ and $Q=50$ give $\gamma=8\times10^{-4}$.  The exact relative propagation-shift error is then $8.0\times10^{-4}$ and $1.6\times10^{-3}$ for the $N=3$ sectors $\eta=+$ and $-$, respectively, and $3.2\times10^{-3}$--$4.0\times10^{-3}$ for $N=9$.  The corresponding phase errors over one diffraction length $z_D$ are about $6.4\times10^{-3}$ rad and $6.4\times10^{-2}$ rad.  For the lowest nonexceptional mode, the same $\gamma$ gives $f_z=1.6\times10^{-3}$ and a transverse-profile infidelity $1-\mathcal F_\perp\simeq1.3\times10^{-6}$.  These magnitudes follow the paraxial ordering: the neglected longitudinal field carries a fraction $O(Q^{-2})$ of the norm, the transverse-profile infidelity is $O(Q^{-4})$, and the modal phase errors accumulated over $z_D$ are $O(Q^{-2})$.

\section{Nonuniform defect density}
\label{sec:graded}

The derivation up to Eq.~\eqref{eq:paraxial} did not assume a particular defect-density profile.  For any smooth axisymmetric density $\sigma(\rho)$, Eq.~\eqref{eq:defect_map} gives
\begin{equation}
f(\rho)=b\int_0^\rho \sigma(\rho')\rho'\,d\rho',\qquad
\Omega(\rho)=\frac{b\sigma(\rho)}{2},
\label{eq:general_sigma_map}
\end{equation}
so the medium is inhomogeneous exactly when the orthonormal quantity $f'(\rho)/\rho=b\sigma(\rho)$ varies across the illuminated region; a factor of $\rho$ in the coordinate coefficient $f'(\rho)$ alone does not imply an inhomogeneous density.  The leading paraxial Hamiltonian derived from Maxwell's equations is
\begin{equation}
H_p=H_{\rm orb}[f]\I-\Omega(\rho)\sigma_3.
\label{eq:Hgraded_general}
\end{equation}
No polarization coupling has been added independently of the metric: the spatial dependence of the helicity term comes from the same density profile that fixes $f(\rho)$.

\subsection{Factorization requires a uniform density}

Equation~\eqref{eq:Hgraded_general} also gives the converse of the result for a uniform density.  If $\Omega=\Omega_g$ is constant, the propagator factorizes as in Eq.~\eqref{eq:factorization}.  Conversely, suppose that every separable input $u(\mathbf r_\perp)\chi$ is to remain separable under the infinitesimal evolution generated by Eq.~\eqref{eq:Hgraded_general}.  The two circular components then evolve with
\begin{equation}
H_\pm=H_{\rm orb}\mp\Omega(\rho).
\end{equation}
For arbitrary $u$, the difference $(H_+-H_-)u=-2\Omega(\rho)u$ can be a spatially independent polarization generator only if $\Omega$ is proportional to the identity on the illuminated connected domain.  Hence
\begin{equation}
\Omega(\rho)=\mathrm{const.}
\iff
U(z)=U_{\rm orb}(z)\otimes U_{\rm pol}(z)
\label{eq:uniformity_iff}
\end{equation}
for arbitrary spatial--polarization inputs.  The equivalence refers to the whole family $U(z)$, or equivalently to its generator near $z=0$; a particular state may still return to a product form at an isolated distance.
For a nonuniform profile, the orbital and helicity parts also fail to commute.  With $\Ag=f(\rho)\hat{\bm\phi}/\rho$ and $\Ag\cdot\nabla\Omega=0$,
\begin{equation}
[H_{\rm orb},\Omega]=-
\frac{1}{2k_0}\left[(\nabla_\perp^2\Omega)+2\nabla_\perp\Omega\cdot\nabla_\perp\right].
\label{eq:comm_grad}
\end{equation}
The gradient of $\Omega(\rho)$, which follows from the metric, is thus what prevents factorization; no additional interaction is involved.

\subsection{Pauli form and zero-shift modes}
\label{sec:pauli_factorization}

The helicity and orbital terms in Eq.~\eqref{eq:Hgraded_general} are not independent.  Let
\begin{equation}
\begin{aligned}
\Pi_j&=-i\partial_j-k_0A_{g,j},\\
\Pi_\pm&=\Pi_x\pm i\Pi_y,\qquad
[\Pi_x,\Pi_y]=ik_0B_g.
\end{aligned}
\label{eq:pi_commutator}
\end{equation}
where $B_g=\partial_xA_{g,y}-\partial_yA_{g,x}=2\Omega(\rho)$.
With Pauli matrices acting on the circular-polarization doublet,
\begin{align}
H_p&=\frac{1}{2k_0}(\sigma_1\Pi_x+\sigma_2\Pi_y)^2\nonumber\\
&=\frac{\Pi_x^2+\Pi_y^2}{2k_0}\I-\frac{B_g}{2}\sigma_3
=\frac1{2k_0}
\begin{pmatrix}\Pi_-\Pi_+&0\\0&\Pi_+\Pi_-\end{pmatrix}.
\label{eq:pauli_factorization}
\end{align}
The sign is fixed by Eq.~\eqref{eq:pi_commutator}.  This is an identity for the paraxial Hamiltonian already derived; it does not introduce an electron-spin or constitutive coupling.  On the infinite plane, with the physical self-adjoint domain, it implies $\langle\Psi,H_p\Psi\rangle\ge0$.  The two polarization partners have the same nonzero eigenvalues when the first-order operators map normalizable states into each other, while zero-shift states can occur in only one of them, as in the Pauli and Aharonov--Casher problem~\cite{AharonovCasher1979,CooperKhareSukhatme1995}.

For $\Omega$ positive and asymptotically bounded below by a positive constant, the regular solutions of $\Pi_+u_{m,+}=0$ are
\begin{equation}
\begin{aligned}
u_{m,+}(\rho,\phi)&=\mathcal N_m\rho^m e^{im\phi}\\
&\quad\times\exp\!\left[-k_0\int_0^\rho ds\,\frac{f(s)}{s}\right],
\quad m\ge0.
\end{aligned}
\label{eq:pauli_zero_mode}
\end{equation}
The positive asymptotic density makes these states square integrable; for a negative asymptotic field the roles of the two circular components reverse.  If the total geometric flux is finite, normalizability restricts the admissible angular indices; a finite aperture changes the boundary domain and need not preserve zero shifts.  Here zero shift means $\beta=0$ relative to the carrier $k_0$, not zero optical frequency.  Appendix~\ref{app:pauli_modes} derives the first-order equation and shows that its regular solutions belong to the transverse exceptional Maxwell sector, with $k_z=k_0$ and $E_z=0$.

\subsection{Polarization texture for an arbitrary density}

Consider an initially linearly polarized separable beam with transverse intensity $I_0(\mathbf r_\perp)=|u_0|^2$ and uniform polarization angle $\theta_0$.  Introduce the normalized input measure
\begin{equation}
d\mu_0=\frac{I_0(\mathbf r_\perp)d^2r}{P_0},\qquad
P_0=\int d^2r\,I_0,
\label{eq:mu0}
\end{equation}
Define the normalized Stokes fields and their intensity-weighted transverse averages by
\begin{equation}
s_j=\frac{S_j}{S_0},\qquad
\bar s_j=\int s_j\,d\mu_0,\qquad j=1,2,
\end{equation}
and the linear-Stokes texture measure
\begin{equation}
\mathcal C^2=
\int\left[(s_1-\bar s_1)^2+(s_2-\bar s_2)^2\right]d\mu_0.
\label{eq:C}
\end{equation}
At $z=0$, this measure uses the input weight.  In the numerical calculations below, the same definition is evaluated with the intensity weight at distance $z$; this affects only terms beyond the leading short-distance coefficient derived here.  The same measure was defined in Ref.~\cite{SilvaPRA2026}; the definition does not depend on the derivation discussed in Sec.~\ref{sec:relation_prior}.
The beam-weighted characteristic function of the dislocation density is
\begin{equation}
\Phi_\sigma(t)=\int e^{-it\sigma(\mathbf r_\perp)}d\mu_0.
\label{eq:char_sigma}
\end{equation}
At short propagation distance, the full paraxial equations give
\begin{align}
s_1+i s_2
&=e^{2i\theta_0}e^{-2i\Omega(\mathbf r_\perp)z}+O(z^2)\nonumber\\
&=e^{2i\theta_0}e^{-ibz\sigma(\mathbf r_\perp)}+O(z^2).
\label{eq:general_local_stokes}
\end{align}
In the local-phase approximation, which keeps this helicity phase and neglects transverse redistribution of the field, averaging Eq.~\eqref{eq:general_local_stokes} gives
\begin{equation}
\left\langle s_1+i s_2\right\rangle_{\rm loc}
=e^{2i\theta_0}\Phi_\sigma(bz).
\label{eq:characteristic_theorem}
\end{equation}
The polarization coherence is therefore the characteristic function of the defect density, with the intensity of the incident beam as the probability weight.

With the texture measure defined by Eq.~\eqref{eq:C}, the same result gives
\begin{equation}
\mathcal C_{\rm loc}^2(z)=1-|\Phi_\sigma(bz)|^2.
\label{eq:C_characteristic}
\end{equation}
Let $\kappa_n[\sigma]$ denote the cumulants with respect to $d\mu_0$.  Since
\begin{equation}
\ln\Phi_\sigma(bz)=\sum_{n\ge1}\frac{(-ibz)^n}{n!}\kappa_n[\sigma],
\label{eq:cumulant_generator}
\end{equation}
we obtain, whenever $\Phi_\sigma\neq0$, for the mean polarization angle $\bar\psi_{\rm loc}\equiv\frac12\arg\langle s_1+is_2\rangle_{\rm loc}$ and for the texture measure
\begin{align}
\bar\psi_{\rm loc}(z)
&=\theta_0-\frac{b}{2}\kappa_1[\sigma]z
+\frac{b^3}{12}\kappa_3[\sigma]z^3+O(z^5),
\label{eq:mean_angle_cumulants}\\
\ln\!\left[1-\mathcal C_{\rm loc}^2(z)\right]
&=-b^2\kappa_2[\sigma]z^2
+\frac{b^4}{12}\kappa_4[\sigma]z^4+O(z^6).
\label{eq:texture_cumulants}
\end{align}
The first cumulant gives the mean geometric rotation, and the variance gives the initial loss of polarization uniformity across the beam.  Expanding the full leading-paraxial evolution, not only the local phase, gives the universal onset
\begin{equation}
\mathcal C(z)=|b|\,|z|\sqrt{\operatorname{Var}_0[\sigma]}+O(z^2)
\label{eq:C_sigma_general}
\end{equation}
Equivalently, $\mathcal C(z)=2|z|\sqrt{\operatorname{Var}_0[\Omega]}+O(z^2)$.
In the local-phase approximation, the next correction is $O(|z|^3)$.  To first order, texture appears if and only if the incident beam samples a nonzero variance of the defect density.  For a continuous density on a connected illuminated region, zero variance means that $\sigma$ is constant there.

Equation~\eqref{eq:C_sigma_general} holds at leading paraxial order.  Appendix~\ref{app:paraxial} also gives the order of magnitude of the field terms dropped in the paraxial reduction.  If $Q=k_0w\gg1$ and $L_g$ is the transverse scale on which $\Omega$ varies, their relative size is, schematically,
\begin{equation}
\epsilon_{\rm vec}=O(Q^{-2})+O\!\left(Q^{-2}\frac{w}{L_g}\right)
+O\!\left(\frac{|\Omega|}{k_0}\right),
\label{eq:graded_vector_error}
\end{equation}
with the last term also scaling as $O(Q^{-2})$ at fixed beam-scale $\Gamma=\Omega k_0w^2$.  These estimates are taken relative to the transverse kinetic term.  The local helicity term is not a suitable reference because it vanishes wherever $\Omega=0$; at such points the gradient contribution must be compared with the covariant transverse scale of Eq.~\eqref{eq:covariant_paraxial_ordering}.  The power counting supports the leading paraxial result, but it does not prove that the coefficient in Eq.~\eqref{eq:C_sigma_general} receives only a multiplicative correction $O(\epsilon_{\rm vec})$ in the full graded Maxwell problem.  That would require the next vector order for a nonuniform profile, which we do not attempt here.  We therefore use Eq.~\eqref{eq:C_sigma_general} only at leading paraxial order.  The exact Maxwell calculations of Secs.~\ref{sec:maxwell_paraxial_validity} and \ref{sec:graded_exact} test propagation constants independently, but they do not replace a nonparaxial derivation of the texture coefficient for a graded density.

\subsection{Local circular polarization and conservation of the integrated \texorpdfstring{$S_3$}{S3}}

The circular sectors in Eq.~\eqref{eq:Hgraded_general} are uncoupled, and each is generated by a Hermitian Hamiltonian.  Their total powers are separately conserved.  For an initially linear state, $P_+(0)=P_-(0)$, so
\begin{equation}
\int d^2r\,S_3(\mathbf r_\perp,z)=0
\quad\text{for all }z
\label{eq:global_S3_zero}
\end{equation}
within the lossless leading-paraxial theory.  We have not shown that this conservation law holds for the full vector Maxwell field, since the transverse $S_3$ omits the longitudinal field components and the vector corrections.  A graded geometry can nevertheless redistribute helicity locally.  The second-order result below assumes a beam with an initially uniform linear Jones state, a real axisymmetric spatial envelope, a smooth axisymmetric defect density, and no initial radial probability current.  For this class of inputs, the continuity equations give
\begin{equation}
\frac{S_3}{S_0}=-\frac{z^2}{2k_0I_0}
\nabla_\perp\cdot\left(I_0\nabla_\perp\Omega\right)+O(z^4)
\label{eq:S3_sigma_general}
\end{equation}
Using Eq.~\eqref{eq:general_sigma_map}, this is equivalently
\begin{equation}
\frac{S_3}{S_0}=-\frac{b z^2}{4k_0I_0}
\nabla_\perp\cdot\left(I_0\nabla_\perp\sigma\right)+O(z^4).
\end{equation}
The leading contribution integrates to zero by the divergence theorem for a localized beam, consistent with Eq.~\eqref{eq:global_S3_zero}.  Equations~\eqref{eq:C_sigma_general} and \eqref{eq:S3_sigma_general} therefore describe effects of different order: density variance produces spatially varying linear polarization at $O(z)$, whereas local circular polarization, for the real axisymmetric inputs considered, requires both the density gradient and transverse transport and appears at $O(z^2)$.  Complex or azimuthally structured inputs can behave differently at short distances.

\subsection{Polarization coherence and spatial--polarization nonseparability}
\label{sec:nonseparability}

The Stokes texture can also be described in mode space.  For a finite-power, fully coherent transverse Jones beam, define the polarization-reduced matrix
\begin{equation}
\begin{aligned}
\rho_{\rm pol}(z)&=\frac{1}{P_0}\int d^2r_\perp\,
\Psi(\mathbf r_\perp,z)\Psi^\dagger(\mathbf r_\perp,z),\\
P_0&=\int S_0\,d^2r_\perp.
\end{aligned}
\label{eq:pol_reduced}
\end{equation}
The instantaneous normalized intensity measure is $d\mu_z=S_0(\mathbf r_\perp,z)d^2r_\perp/P_0$ and we write $\bar s_j=\int s_jd\mu_z$.  The matrix $\rho_{\rm pol}$ is positive and has unit trace; its two eigenvalues and linear entropy are
\begin{align}
\lambda_\pm&=\frac{1\pm\sqrt{\bar s_1^2+\bar s_2^2+\bar s_3^2}}{2},\nonumber\\
\mathcal L(z)&\equiv1-\operatorname{Tr}\rho_{\rm pol}^2
=\frac{1-\bar s_1^2-\bar s_2^2-\bar s_3^2}{2}.
\label{eq:pol_linear_entropy}
\end{align}
At every illuminated point, the fully coherent Jones field obeys $s_1^2+s_2^2+s_3^2=1$.  Combining this identity with Eq.~\eqref{eq:C}, evaluated with $d\mu_z$, gives the relation
\begin{equation}
2\mathcal L=\mathcal C^2+\langle s_3^2\rangle_z-\bar s_3^2.
\label{eq:texture_entropy_exact}
\end{equation}
which is exact within the leading paraxial theory.  For an initially linearly polarized beam, separate conservation of circular powers in Eq.~\eqref{eq:global_S3_zero} implies $\bar s_3=0$ at all $z$, so
\begin{equation}
\mathcal L=\tfrac12\left[\mathcal C^2+\langle s_3^2\rangle_z\right].
\label{eq:texture_entropy_linear}
\end{equation}
In the local-phase approximation $s_3\equiv0$ and Eq.~\eqref{eq:characteristic_theorem} yields
\begin{equation}
\lambda_\pm^{\rm loc}=\frac{1\pm|\Phi_\sigma(bz)|}{2},\qquad
\mathcal L_{\rm loc}=\frac{1-|\Phi_\sigma(bz)|^2}{2}
=\frac{\mathcal C_{\rm loc}^2}{2}.
\label{eq:entropy_characteristic}
\end{equation}
Consequently, $\mathcal L_{\rm loc}=b^2z^2\operatorname{Var}_0[\sigma]/2+O(z^4)$.  In the full propagation, local circular polarization adds the term $\langle s_3^2\rangle_z/2$, which begins at order $z^4$ under the assumptions of Eq.~\eqref{eq:S3_sigma_general}.  Equations~\eqref{eq:texture_entropy_exact}--\eqref{eq:entropy_characteristic} connect the measured Stokes statistics with the classical spatial--polarization nonseparability of vector beams~\cite{QianEberly2011,McLaren2015}; they say nothing about quantum entanglement of an ordinary laser field.  A finite set of measured cumulants also does not determine the defect profile uniquely; imaging the profile would require spatial scanning or several input profiles.

\subsection{A saturating density profile}

As an example that is regular everywhere, take the density
\begin{equation}
\sigma(\rho)=\sigma_0+\Delta\sigma\left(1-e^{-\rho^2/R^2}\right),
\label{eq:sigma_sat}
\end{equation}
which is regular on the axis and approaches the finite asymptotic value $\sigma_0+\Delta\sigma$.  No optical potential is specified independently.  Using Eqs.~\eqref{eq:general_sigma_map} and \eqref{eq:Omega}, define
\begin{equation}
\Omega_0=\frac{b\sigma_0}{2},\qquad
\Delta\Omega=\frac{b\Delta\sigma}{2}.
\end{equation}
Then
\begin{equation}
\Omega(\rho)=\Omega_0+\Delta\Omega\left(1-e^{-\rho^2/R^2}\right),
\label{eq:Omega_sat}
\end{equation}
and direct integration of Eq.~\eqref{eq:general_sigma_map} gives
\begin{equation}
f(\rho)=(\Omega_0+\Delta\Omega)\rho^2
-\Delta\Omega R^2\left(1-e^{-\rho^2/R^2}\right).
\label{eq:f_sat}
\end{equation}
The optical problem is thus still fixed by the defect density, the resulting metric, and minimally coupled Maxwell theory.

Introduce
\begin{equation}
r=\frac{\rho}{w_0},\qquad
\zeta=\frac{z}{z_D},\qquad z_D=k_0w_0^2,
\end{equation}
and
\begin{equation}
\Gamma_0=\Omega_0z_D,\qquad
\Delta\Gamma=\Delta\Omega z_D,\qquad
s=\frac{R}{w_0}.
\label{eq:sat_controls}
\end{equation}
The dimensionless local twist and the dimensionless azimuthal component of the geometric potential are
\begin{align}
V_s(r)&=\Gamma_0+\Delta\Gamma\left(1-e^{-r^2/s^2}\right),
\label{eq:Vsat}\\
a_\phi(r)&=(\Gamma_0+\Delta\Gamma)r
-\frac{\Delta\Gamma s^2}{r}\left(1-e^{-r^2/s^2}\right).
\label{eq:asat}
\end{align}
The apparent $1/r$ is removable:
\begin{equation}
a_\phi(r)=\Gamma_0r+\frac{\Delta\Gamma}{2s^2}r^3+O(r^5),
\end{equation}
while
\begin{equation}
V_s(r)=\Gamma_0+\frac{\Delta\Gamma}{s^2}r^2+O(r^4).
\label{eq:sat_local_quad}
\end{equation}
A purely quadratic density profile is only the near-axis expansion of this saturating profile.

\begin{figure}[t]
\includegraphics[width=\columnwidth]{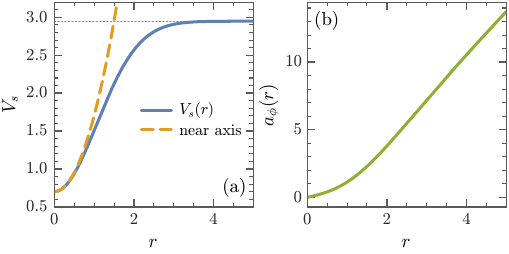}
\caption{Saturating defect-density profile of Eq.~\eqref{eq:sigma_sat} for $\Gamma_0=0.7$, $\Delta\Gamma=2.25$, and $s=1.5$.  (a) $V_s=z_D\Omega$ approaches a finite asymptote (dotted line); the dashed curve is its near-axis quadratic approximation.  (b) Azimuthal coefficient $a_\phi$ obtained from the same metric function $f(\rho)$.  The density and the twist are bounded, whereas $a_\phi$ grows linearly at large $r$ because the density tends to a nonzero constant.}
\label{fig:sat_profile}
\end{figure}

Let $\epsilon=(k_0w_0)^{-1}\ll1$.  In these dimensionless variables, the transverse condition invariant under the coordinate gauge and the remaining sufficient conditions are
\begin{align}
\epsilon_{\rm cov}
&=\epsilon\,
\frac{\|[\nabla_r-i a_\phi(r)\hat{\bm\phi}]\Psi\|}{\|\Psi\|}\ll1,\nonumber\\
\frac{|\Omega|}{k_0}&=\epsilon^2|V_s(r)|\ll1,\nonumber\\
\frac{|\partial_\rho\Omega|}{k_0^2}
&=2|\Delta\Gamma|\epsilon^3\frac{r}{s^2}e^{-r^2/s^2}\ll1.
\label{eq:sat_validity}
\end{align}
In the regular axisymmetric coordinate gauge, $\epsilon\max\{1,|a_\phi|\}\ll1$ over the beam core is a convenient sufficient condition; unlike $\epsilon_{\rm cov}$, it depends on that gauge choice.
The paraxial equation is then
\begin{equation}
 i\partial_\zeta\Psi=
 \left[-\frac12\left(\nabla_r-i a_\phi(r)\hat{\bm\phi}\right)^2\I
 -V_s(r)\sigma_3\right]\Psi,
\label{eq:sat_dimless}
\end{equation}
and in the axisymmetric orbital sector
\begin{equation}
 i\partial_\zeta\psi_h=
 \left[-\frac12(\partial_r^2+r^{-1}\partial_r)
 +\frac12a_\phi(r)^2-hV_s(r)\right]\psi_h.
\label{eq:sat_radial}
\end{equation}
Figure~\ref{fig:sat_profile} shows the saturating twist profile and the azimuthal coefficient derived from it.

For $|u_0|^2\propto e^{-2r^2}$, set $X=e^{-r^2/s^2}$.  Direct integration gives
\begin{equation}
\langle X\rangle_0=\frac{2s^2}{2s^2+1},\qquad
\langle X^2\rangle_0=\frac{s^2}{s^2+1},
\end{equation}
and hence
\begin{equation}
\operatorname{Var}_0[V_s]
=(\Delta\Gamma)^2\frac{s^2}{(s^2+1)(2s^2+1)^2}.
\label{eq:var_sat}
\end{equation}
Equation~\eqref{eq:C_sigma_general} becomes
\begin{equation}
\mathcal C(\zeta)=
\frac{2|\Delta\Gamma|s}{(2s^2+1)\sqrt{s^2+1}}|\zeta|
+O(\zeta^2).
\label{eq:C_sat_slope}
\end{equation}
For this profile Eq.~\eqref{eq:characteristic_theorem} can be evaluated in closed form~\cite{NISTHandbook}:
\begin{align}
\left\langle s_1+i s_2\right\rangle_{0,\rm loc}
={}&e^{-2i(\Gamma_0+\Delta\Gamma)\zeta}
\,{}_1F_1\!\left(2s^2;2s^2+1;2i\Delta\Gamma\zeta\right),
\label{eq:sat_localphase_average}
\end{align}
so that
\begin{equation}
\mathcal C_{\rm loc}(\zeta)=
\left[1-\left|{}_1F_1\!\left(2s^2;2s^2+1;2i\Delta\Gamma\zeta\right)\right|^2\right]^{1/2}.
\label{eq:C_sat_exactlocal}
\end{equation}
For the same Gaussian input, Eq.~\eqref{eq:S3_sigma_general} gives
\begin{equation}
\frac{S_3}{S_0}=
-\frac{2\Delta\Gamma}{s^2}\zeta^2 e^{-r^2/s^2}
\left[1-\left(2+\frac{1}{s^2}\right)r^2\right]
+O(\zeta^4).
\label{eq:S3_sat}
\end{equation}

\subsection{Numerical results}

\begin{figure}[!tbp]
\includegraphics[width=\columnwidth]{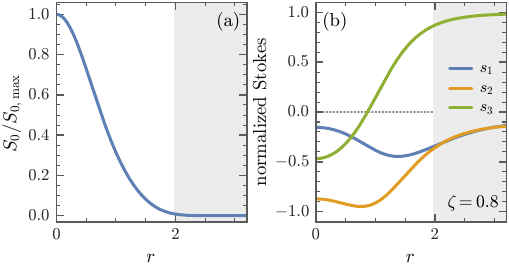}
\caption{Stokes profiles for the saturating density of Eq.~\eqref{eq:sigma_sat} with $\Gamma_0=0.7$, $\Delta\Gamma=2.25$, and $s=1.5$, at $\zeta=0.8$: (a) intensity and (b) normalized Stokes parameters.  The linear Stokes components vary across the beam, and diffraction produces a nonzero local $S_3$, although the density is smooth and bounded.  In the shaded region $S_0/S_{0,\max}<10^{-2}$; the normalized Stokes parameters are still defined there but carry negligible power.}
\label{fig:sat_stokes}
\end{figure}

We solve Eq.~\eqref{eq:sat_radial} for an initially linearly polarized Gaussian by diagonalizing a finite-difference form of each radial operator and propagating with its exact spectral phases (Appendix~\ref{app:numerics}).  Unless stated otherwise, we use
\begin{equation}
\Gamma_0=0.7,\qquad s=1.5,\qquad \Delta\Gamma=2.25,
\end{equation}
so that the near-axis curvature $\Delta\Gamma/s^2=1$.  Figure~\ref{fig:sat_stokes} shows the resulting Stokes structure at $\zeta=0.8$.

\begin{figure}[!tbp]
\includegraphics[width=\columnwidth]{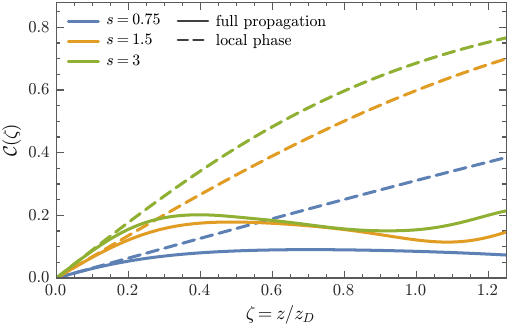}
\caption{Texture measure for saturating defect profiles with fixed $\Gamma_0=0.7$ and $\Delta\Gamma/s^2=1$.  Solid curves are full leading-paraxial propagation; dashed curves are the exact local-phase expression in Eq.~\eqref{eq:C_sat_exactlocal}.}
\label{fig:sat_C}
\end{figure}

Figure~\ref{fig:sat_C} compares three saturation radii while keeping the same near-axis curvature.  The dashed curves are Eq.~\eqref{eq:C_sat_exactlocal}; the solid curves are the full leading-paraxial propagation.

\begin{figure}[!tbp]
\includegraphics[width=\columnwidth]{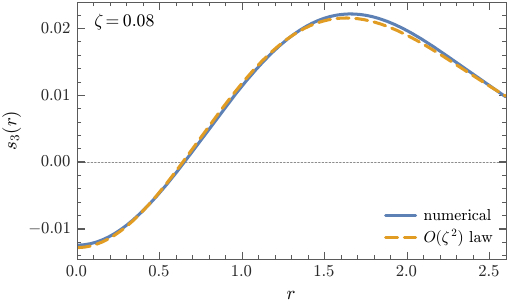}
\caption{Local circular polarization produced by diffraction in the saturating profile at $\zeta=0.08$.  The numerical $s_3=S_3/S_0$ agrees with Eq.~\eqref{eq:S3_sat} across the beam core.}
\label{fig:sat_s3}
\end{figure}

Figure~\ref{fig:sat_s3} compares the second-order law for the circular polarization with the numerical result.

\begin{figure}[!tbp]
\includegraphics[width=\columnwidth]{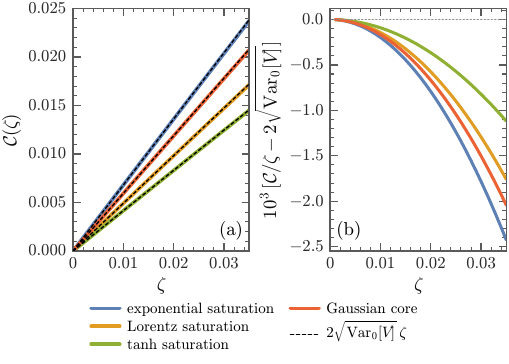}
\caption{Initial texture growth for different density profiles.  (a) Colored curves are full leading-paraxial propagation for the four smooth bounded defect-density profiles of Appendix~\ref{app:numerics}; black dashed lines are the universal predictions $\mathcal C=2\sqrt{\operatorname{Var}_0[V]}|\zeta|$.  (b) Residual $\mathcal C/\zeta-2\sqrt{\operatorname{Var}_0[V]}$ of the same calculations, which vanishes quadratically as $\zeta\to0$.  The short-distance agreement shows that the onset is set by the beam-weighted density variance, whatever the shape of the profile.}
\label{fig:general_universality}
\end{figure}

To test Eq.~\eqref{eq:C_sigma_general} beyond the analytically solvable profile, we also propagate four smooth bounded density profiles: the exponential saturation above, a Lorentz-type saturation, a $\tanh$ saturation, and a Gaussian core excess, listed in Appendix~\ref{app:numerics}.  Each profile enters only through $\sigma(\rho)$, that is, through Eq.~\eqref{eq:general_sigma_map}.  Figure~\ref{fig:general_universality} compares the full numerical texture with the parameter-free slopes $2\sqrt{\operatorname{Var}_0[V]}$.  At $\zeta=0.002$, the numerical values of $\mathcal C/\zeta$ differ from the continuum variance predictions by less than $8\times10^{-6}$ for all four tests.  This small offset is the next term of the short-distance expansion, not a discretization error.  For real axisymmetric inputs, the reduced radial operators are real, so $\mathcal C^2(z)$ is even in $z$ and the remainder in Eq.~\eqref{eq:C_sigma_general} is $O(|z|^3)$.  Accordingly, the residual $\mathcal C/\zeta-2\sqrt{\operatorname{Var}_0[V]}$ scales as $\zeta^2$ [Fig.~\ref{fig:general_universality}(b)], and a two-point Richardson extrapolation~\cite{Richardson1911} from $\zeta=0.002$ and $0.004$ reproduces the predicted slopes to within $1.3\times10^{-6}$.  A separate calculation in a Laguerre spectral basis gives the same offsets and reduces this difference below $10^{-9}$ (Appendix~\ref{app:numerics}).  The integrated normalized $S_3$ remains below $10^{-14}$ in magnitude, i.e., at machine precision.

At leading order, the conclusion does not depend on the profile: a uniform density produces only a common geometric rotation of the polarization, whereas any density variation sampled by the beam produces linear-Stokes texture from the start.  Diffraction then produces local circular polarization one order later, with the integrated $S_3$ unchanged.

\subsection{Exact Maxwell modes for a graded density}
\label{sec:graded_exact}

The modal Maxwell equation can also be solved without setting $\Omega'$ to zero.  In each Beltrami sector $\mathcal C_f\mathbf E_\eta=\eta k_0\mathbf E_\eta$, its longitudinal component satisfies the exact radial equation of Appendix~\ref{app:graded_maxwell}, with the full $f(\rho)$ determined by Eq.~\eqref{eq:general_sigma_map}.  For the saturating profile of Eq.~\eqref{eq:sigma_sat}, we compare the forward Maxwell roots $k_{z,\eta}$, obtained from a separate radial eigenvalue calculation, with the leading paraxial shifts $\beta_{p,\eta}$ at $Q=k_0w_0=50$, $100$, and $200$.

For the first four nonzero positive-helicity modes with longitudinal indices $m=1,2$ the largest absolute relative shift errors are $1.14\times10^{-3}$, $2.84\times10^{-4}$, and $7.09\times10^{-5}$, respectively.  For the corresponding four physical negative-helicity modes at longitudinal $m=0$, the maxima are $2.67\times10^{-3}$, $6.66\times10^{-4}$, and $1.66\times10^{-4}$.  The errors decrease as $Q^{-2}$ for fixed low-lying modes, as expected; the normalizable transverse zero-shift family is treated separately and has $\beta=0$ already in exact Maxwell theory.  Appendix~\ref{app:graded_maxwell} gives the radial eigenproblem, the mode labels, the numerical method, and the convergence tests.  These comparisons test propagation constants for this profile; they do not give a general full-vector bound on the correction to the short-distance texture coefficient in graded media, which remains an open problem.

\section{Discussion}
\label{sec:discussion}

Four effects should be kept apart.  First, geometric optics gives a rotation of the transverse basis beyond eikonal order, while the light cone is the same for both polarizations.  Second, a beam of finite width probes the geometry through the orbital gauge field $\Ag$, which turns the envelope equation into a Landau problem.  Third, because the helicity term is spatially constant for a uniform density, the leading paraxial propagator factorizes and cannot make a separable input nonseparable.  Fourth, a gradient of the defect density makes the polarization rotation depend on position and, together with diffraction, produces Stokes textures.

The kernel~\eqref{eq:Landau_kernel} gives access to observables beyond those of centered Gaussian beams.  The simplest is the rotation of $\pm\ell$ petal patterns, Eq.~\eqref{eq:petal_rate}: the orbital pattern rotates at the same rate as a uniform linear polarization, so the angle between them stays fixed during propagation.  This co-rotation is geometric and is not spin--orbit conversion, because the OAM weights and the spatial--polarization Schmidt spectrum are unchanged.  The screw geometry thus combines great changes of the orbital structure with polarization transport without making the two degrees of freedom nonseparable.

The general density law also suggests a way to probe defect statistics optically.  In the metric-Maxwell model, a beam scanned across the sample samples a different measure $d\mu_0$ at each position: its mean polarization rotation gives $\langle\sigma\rangle_0$, its initial texture growth gives $\sqrt{\operatorname{Var}_0[\sigma]}$, and higher local-phase cumulants carry further information.  These observables do not by themselves determine $\sigma(\rho)$ uniquely; the numerical aperture, the interface Jones matrices, any material birefringence, and the weighting function must be characterized separately.  The nonseparability relation~\eqref{eq:texture_entropy_linear} provides a second quantity, based on polarization coherence, for such measurements.  A system that reproduces only the orbital Hamiltonian, such as the rotating graded-index analogue below, need not reproduce the polarization or interface response of the metric Maxwell system.

The Landau mapping links the continuum-defect problem to the literature on synthetic magnetic fields for light~\cite{Schine2016,Guglielmon2021,Barczyk2024,Barsukova2024,Song2025}; Landau quantization itself is not new.  What differs is the origin of the effective vector potential: it is not produced by resonator design or by a strain pattern in a photonic crystal, but follows from the screw-dislocation metric through the dual frame and the paraxial reduction of the vector Maxwell equations.  Within the stated approximation, the matched-Gaussian condition of Eq.~\eqref{eq:stationary}, the OAM co-rotation, and the Landau-mode spacing depend only on $\Omega_g$.  The exact Maxwell comparisons of Secs.~\ref{sec:maxwell_paraxial_validity} and \ref{sec:graded_exact} also show how the accuracy depends on the mode: at fixed $\Gamma$, higher Landau indices need larger $Q=k_0w_0$ to keep both the propagation-shift error and the accumulated phase error small.  This refines the usual condition $Q\gg1$.

\subsection{Relation to earlier work on the same geometry}
\label{sec:relation_prior}

The metric of Eq.~\eqref{eq:metric} has appeared in several earlier papers by the author, so we state which of their conclusions the present derivation changes.  Sharing the geometry does not mean sharing the same error: the assumptions involved differ from paper to paper, and conclusions that do not depend on them are not affected.

Reference~\cite{SilvaPRA2026} is affected most directly.  The radius-dependent optical splitting was obtained after combining a torsionful Maxwell rewriting with an inconsistent truncation of the connection terms and then reading the coordinate factor in $K_{\rho\phi z}=\Omega_g\rho$ as a local physical strength.  The present orthonormal-frame calculation instead gives $K_{\hat\rho\hat\phi\hat z}=\Omega_g$ for a uniform density, while minimal Maxwell propagation is governed by the metric and its Levi--Civita connection.  Locally, optical transport in a uniform density is the same everywhere, and the radial Stokes texture reported in Ref.~\cite{SilvaPRA2026} does not survive the corrected derivation.  In magnitude, the splitting $\Delta k_z=2\Omega_o\rho=2k_0\Omega_g\rho$ reported there exceeds the $\rho$-independent separation $2|\Omega_g|$ between the circular propagation constants of Eq.~\eqref{eq:paraxial} at fixed transverse profile by the factor $k_0\rho$, which is about $59$ for $\rho=10~\mu$m and $\lambda=1.064~\mu$m.

Reference~\cite{SilvaPRA2026} also set aside the quadratic term $(\Omega_o\rho)^2/(2k_0)$, common to both of its local circular propagation constants, as a global phase.  Written with $\Omega_o=k_0\Omega_g$, this term has the magnitude of the confinement term $k_0\Omega_g^2\rho^2/2$ in Eq.~\eqref{eq:Huniform}, but it enters the effective potential of that reduction with the opposite sign, and the rotation term $-\Omega_gL_z$ is absent.  The breathing of a Gaussian beam in Fig.~\ref{fig:breathing} and the petal rotation in Fig.~\ref{fig:petals} thus have no counterpart there.

Section~VII of Ref.~\cite{SilvaPRA2026} placed the same radial splitting in an externally imposed, q-plate-like polarization frame~\cite{Marrucci2006}.  The selection rule $m\to m\pm q$ found there follows from the imposed frame alone, but the sideband amplitudes and the maps of its Figs.~1 and 2(b)--2(d) were computed with the radius-dependent splitting and are not supported by the present derivation.  For any axisymmetric density, the Hamiltonian~\eqref{eq:Hgraded_general} commutes with $L_z$ and $\sigma_3$, so the OAM weights of each circular component are conserved, in agreement with the $q=0$ spectrum of Fig.~2(a) of Ref.~\cite{SilvaPRA2026}.

The situation is different for Ref.~\cite{SilvaOQE2026}.  Its main quantum Hamiltonian was constructed from the torsion-free Levi--Civita Laplace--Beltrami operator, so the coordinate-contortion problem above does not apply.  The bulk operator and its Landau-like radial confinement remain valid consequences of the metric.  Its finite hard-wall version, however, needs a separate treatment of the boundary.  For the one-dimensional hard-wall problem, the nondegenerate normalized Dirichlet eigenfunction $\chi_n(z)$ may be chosen real (up to a constant phase).  In that standard eigenbasis,
\begin{equation}
\left\langle\chi_n\middle|\partial_z\middle|\chi_n\right\rangle
=\int_{-L/2}^{L/2}\chi_n\,\partial_z\chi_n\,dz
=\frac12\left[\chi_n(z)^2\right]_{-L/2}^{L/2}=0.
\label{eq:hardwall_derivative_zero}
\end{equation}
Hence a term linear in $\partial_z$ has no diagonal first-order contribution within a single hard-wall subband.  The same holds at every order: the Laplace--Beltrami operator of the spatial metric and the Dirichlet conditions at $z=\pm L/2$ are real, so complex conjugation maps the sector $m$ onto $-m$ at the same energy, and $E(m)=E(-m)$ exactly for the finite dot.  The linear $m\Omega k_z$ splitting appears only at fixed $k_z$, in an open or periodic wire, where $E(m,k_z)=E(-m,-k_z)$.  The optical part of that work also needs to be reanalyzed, whereas the metric itself and the translational Landau structure are not affected by the present correction.

We list these points for completeness; formal corrections, where needed, belong to the individual articles.  The present paper provides a derivation from first principles against which the earlier claims can be checked.

\subsection{A rotating graded-index analogue}

A laboratory analogue of the uniform paraxial Hamiltonian can be derived without adding the Landau terms by hand.  Consider an isotropic, nonmagnetic graded-index host with vacuum wavenumber $k_{\rm vac}=\omega/c$, on-axis index $n_0$, carrier wavenumber $k_h=n_0k_{\rm vac}$, and near-axis profile
\begin{equation}
n^2(\rho)=n_0^2\left(1-\alpha^2\rho^2\right)+O(\rho^4).
\label{eq:grin_profile}
\end{equation}
Neglecting material birefringence and the usual higher-order vector corrections, each transverse component obeys the Helmholtz equation $[\nabla_\perp^2+\partial_z^2+k_{\rm vac}^2n^2(\rho)]\Psi=0$.  Writing $\Psi=e^{ik_hz}\psi$ and dropping $\partial_z^2\psi$ at paraxial order gives
\begin{equation}
i\partial_z\psi=\left[-\frac{\nabla_\perp^2}{2k_h}+\frac{k_h\alpha^2\rho^2}{2}\right]\psi.
\label{eq:grin_lab_hamiltonian}
\end{equation}
Now describe the same field in a transverse frame co-rotating at angular rate $\alpha$.  With $U(z)=\exp[-i\alpha z J_z]$, $J_z=L_z+S_z$, and $\psi_{\rm rot}=U^\dagger\psi$, the transformed Hamiltonian is
\begin{align}
H_{\rm rot}
&=U^\dagger H_{\rm lab}U-iU^\dagger\partial_zU\nonumber\\
&=\left[-\frac{\nabla_\perp^2}{2k_h}-\alpha L_z+\frac{k_h\alpha^2\rho^2}{2}\right]\mathbb I-\alpha\sigma_3,
\label{eq:grin_rot_hamiltonian}
\end{align}
where $S_z=\sigma_3$ in the circular convention of Eq.~\eqref{eq:circular}.  Equation~\eqref{eq:grin_rot_hamiltonian} is identical to Eq.~\eqref{eq:Huniform} after $k_h\to k_0$ and $\alpha\to\Omega_g$, so the correspondence follows from the Helmholtz equation and a change of frame.  Its physical meaning is different, however: in the graded-index system the term $-\alpha\sigma_3$ comes from the rotation of the frame and is not by itself a circular birefringence in the laboratory.  Mechanical twist, photoelasticity, intrinsic anisotropy, and interface effects can add material corrections~\cite{SalehTeich2019,Xi2013,Buerger2024}.  A rotating graded-index medium is thus an analogue of the orbital propagator in a rotating frame, not evidence for a screw-dislocation metric.

For orientation, the matched case $|\Gamma|=2$ corresponds to the following laboratory scales.  Taking $\lambda=1.064\,\mu{\rm m}$ in vacuum and $w_0=10\,\mu{\rm m}$ gives $k_0=2\pi/\lambda\simeq5.91\times10^6\,{\rm m}^{-1}$ and $z_D=k_0w_0^2\simeq5.91\times10^{-4}\,{\rm m}$, so the matched condition corresponds to $|\Omega_g|\simeq3.39\times10^3\,{\rm m}^{-1}$.  Within the ideal metric-Maxwell model, this scale would produce a geometric frame rotation $|\Omega_g|L\simeq3.39$ rad over $L=1$ mm.  This is a value of the model, not a prediction for a natural dislocated crystal; applying it to a material requires a separately justified constitutive model (Sec.~\ref{sec:metric_medium}).

The same care is needed in comparisons with twisted fibers and helical waveguides.  Continuous mechanical twist can produce circular birefringence, OAM splitting, optical Zeeman-like shifts, and coordinate-frame-dependent mode formation~\cite{Xi2013,Barshak2015,Buerger2024}.  Observing polarization or OAM rotation is not, by itself, evidence of a screw-dislocation geometry.  What is specific here is the combination of (i) a metric fixed by the defect density, (ii) minimal Maxwell dynamics with no independent chiral constitutive coefficient, and (iii) the resulting factorization theorem for uniform density.  Likewise, the joint rotation of a balanced $\pm\ell$ petal pattern and a uniform linear polarization is not a new twist effect in general; here both rates follow from the same defect geometry while the OAM weights and the spatial--polarization Schmidt spectrum stay unchanged, so no spin--orbit conversion occurs.

The source-driven theory of Ref.~\cite{SilvaAS13014} adds to the same screw geometry an independent gauge-compatible axial electromagnetic response and studies source matching, Beltrami sectors, and the splitting of guided modes.  Here, that constitutive term is zero, and we study finite-beam propagation governed by the metric Maxwell system alone.  The confluent-hypergeometric and Laguerre mode structure and the general Landau correspondence are not claimed as new.  The new elements are the finite-beam propagator for the geometry alone, with its error estimate from the vector Maxwell equations; the factorization theorem for a uniform density; the characteristic-function and cumulant relations for a general density; the onset of texture controlled by the density variance; and the conservation of the integrated $S_3$ with its local generation at $O(z^2)$.

The results for a general density also show which observable measures what.  Equations~\eqref{eq:characteristic_theorem}--\eqref{eq:S3_sigma_general} assign different statistical information about the sampled defect profile to different observables.  In the geometry-adapted bulk frame, the mean rotation rate measures the beam-weighted mean density, and the initial linear-Stokes texture measures its standard deviation.  In the local-phase approximation, the higher odd and even cumulants enter the mean angle and the texture coherence, respectively.  Local circular polarization appears only at second order and is controlled by the divergence of the intensity-weighted density gradient, while the integrated circular Stokes component remains exactly zero.  The numerical tests with four profiles show that this ordering does not depend on the solvable saturating example.

As discussed in Sec.~\ref{sec:metric_medium} and around Eq.~\eqref{eq:interface_jones}, the equations derived here are predictions of the effective metric Maxwell model, not constitutive laws for every crystal containing screw dislocations.  A natural crystal departs from this model through photoelastic and other microscopic responses, whereas a transformation-optics medium can in principle be designed to realize the metric tensors.  This matters especially when polarization rotation is compared with ordinary optical activity.

The geometry-only theory also differs from an effective chiral medium with an independent axial electromagnetic coefficient.  Such a coefficient splits the circular eigenwaves even in a locally homogeneous medium, as in Ref.~\cite{SilvaAS13014}.  Here $\Omega_g$ enters instead through the metric, the transverse orbital connection, and the parallel transport of the frame.  Additional constitutive couplings, finite interfaces, absorption, or nonaxisymmetric defect profiles can add further effects; they are not needed for the orbital effects derived here and must be modeled separately.

\subsection{Proposed polarimetric measurement}
\label{sec:experimental_protocol}

\begin{figure*}[t]
\includegraphics[width=0.94\textwidth]{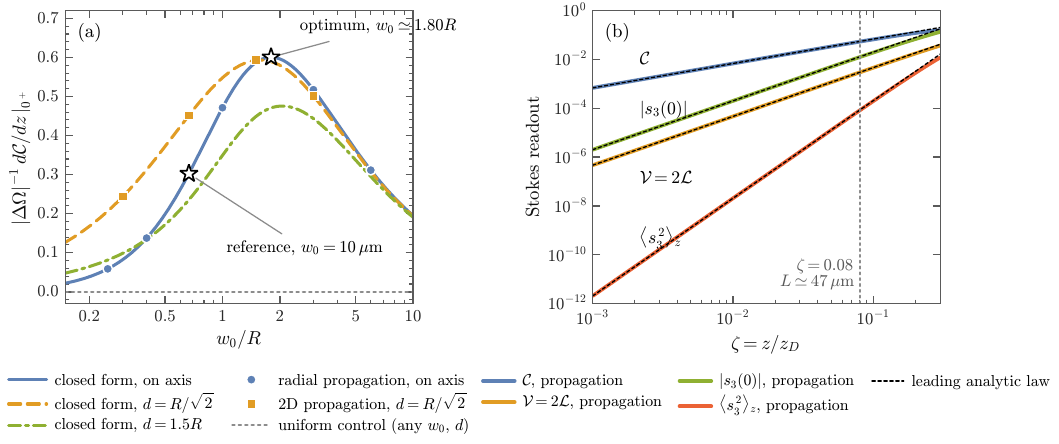}
\caption{Design curves and expected signals for the saturating profile in the ideal metric-Maxwell model.  (a) Initial texture slope per unit length, normalized by $|\Delta\Omega|$, versus input waist for Gaussian beams centered on the defect axis or displaced from it by $d$.  Curves are Eq.~\eqref{eq:waist_slope}; circles are full radial propagation and squares are two-dimensional propagation including the magnetic coupling of the off-axis beam.  Stars mark the reference waist $w_0=10\,\mu$m ($w_0/R=2/3$) and the on-axis optimum $w_0\simeq1.80R$.  A uniform control gives zero for every waist and offset.  (b) Growth with distance for the reference parameters $(\Gamma_0,\Delta\Gamma,s)=(0.7,2.25,1.5)$: full leading-paraxial propagation (solid) of $\mathcal C$, $\mathcal V=2\mathcal L$, the central circularity $|s_3(0)|$, and the circular excess $\langle s_3^2\rangle_z=\mathcal V-\mathcal C^2$, with their leading laws (dashed) $\propto\zeta$, $\zeta^2$, $\zeta^2$, and $\zeta^4$.  The dotted line marks $\zeta=0.08$ ($L\simeq47\,\mu$m).  Interfaces, loss, detection noise, and finite camera resolution are not included.}
\label{fig:experiment_readout}
\end{figure*}

\emph{Sample and measured quantities.} The sample is a slab or guided region whose effective Maxwell response has been shown, independently of the beam measurement, to approximate the impedance-matched constitutive tensors of Eqs.~\eqref{eq:plebanski_metric}--\eqref{eq:metric_medium_lab} over the illuminated aperture and the optical bandwidth.  An ordinary nonmagnetic semiconductor with dislocations does not, in general, have this response.  A transformation-optics medium could realize it in principle; a programmable paraxial emulator that reproduces only Eq.~\eqref{eq:paraxial} can test the envelope predictions, but not the constitutive model or its interfaces.

For a Gaussian beam with calibrated, uniform linear input polarization, an imaging polarimeter measures the six intensity images $I_x,I_y,I_{45},I_{135},I_+,I_-$ used in Eqs.~\eqref{eq:analyzer_intensities}--\eqref{eq:stokes_operational}. The images can be taken with sequential analyzer settings or with calibrated snapshot full-Stokes instruments~\cite{Singh2020,CoxRosales2023}. Exposure and pixel-response calibration, registration between analyzer images, a mask for low-power pixels, and the measured output intensity are needed to evaluate $s_j(\mathbf r_\perp)$ and the instantaneous weight $d\mu_z$. Detection noise biases the squared texture estimators upward: for independent per-pixel errors of variance $\delta_s^2$ in each normalized Stokes component, the estimates of $\mathcal C^2$ and of the variance $\mathcal V$ defined below acquire biases of approximately $2\delta_s^2$ and $3\delta_s^2$. These biases do not decrease with the number of pixels; they must be measured on a reference data set with no expected texture and subtracted, or reduced by binning and frame averaging, and the error analysis must include the correlations between analyzer channels.  Without bias subtraction, measuring the value $\mathcal V\simeq2.9\times10^{-3}$ quoted below to $10\%$ requires $\delta_s\lesssim0.01$. In addition to the linear texture measure $\mathcal C$ of Eq.~\eqref{eq:C}, define the full-Stokes variance
\begin{equation}
\begin{split}
\mathcal V(z)&\equiv\sum_{j=1}^{3}\left(\langle s_j^2\rangle_z-\bar s_j^2\right)\\
&=\mathcal C^2+\langle s_3^2\rangle_z-\bar s_3^2
=2\mathcal L(z).
\end{split}
\label{eq:experimental_variance}
\end{equation}
The last identity follows from Eq.~\eqref{eq:texture_entropy_exact} and expresses $\mathcal L$ through measurable quantities.

The two measures respond differently to uniform, lossless polarization optics.  A Jones unitary acting after the sample (exit face, relay, or imaging optics) rotates every output Stokes vector by the same $SO(3)$ matrix, so $\mathcal V$ is unchanged.  The linear texture $\mathcal C$ is invariant only under rotations about the $S_3$ axis, i.e., under a change of analyzer azimuth, and it changes under residual retardance that mixes $S_3$ with $S_1$ and $S_2$.  A unitary acting before a graded region is different, because it does not commute with the helicity-diagonal evolution generated by Eq.~\eqref{eq:Hgraded_general}.  For an input $\Psi(0)=u\,(a,b)^{T}$ in the circular basis, with $\int|u|^2d^2r_\perp=1$ and $|a|^2+|b|^2=1$, Eq.~\eqref{eq:pol_reduced} gives $\mathcal V=4|a|^2|b|^2(1-|\mathcal O|^2)$, where $\mathcal O(z)=\int(U_-u)^*(U_+u)\,d^2r_\perp$ and $U_h(z)=\exp[-iz(H_{\rm orb}-h\Omega)]$.  Hence
\begin{equation}
\mathcal V=\left(1-\bar s_{3,\rm in}^2\right)\mathcal V_{\rm lin},
\qquad \bar s_{3,\rm in}=|a|^2-|b|^2,
\label{eq:entrance_factor}
\end{equation}
where $\mathcal V_{\rm lin}=1-|\mathcal O|^2$ is the value for a linearly polarized input.  Entrance rotations about the $S_3$ axis leave $\mathcal V$ unchanged; any other entrance retardance rescales it by the factor in Eq.~\eqref{eq:entrance_factor}, which is fixed by the reference measurement of the input state.  In particular, a circularly polarized input gives $\mathcal V=0$ even in a graded region.  For a uniform density, the propagator is a product, so $\mathcal V=0$ for every uniformly polarized input and every uniform entrance or exit unitary.  Spatially varying Jones response, diattenuation, absorption, and depolarization invalidate these statements; they must be characterized or treated as systematic errors.  In the local-phase approximation with a linearly polarized input, Eqs.~\eqref{eq:C_characteristic} and \eqref{eq:entropy_characteristic} imply
\begin{equation}
\mathcal V_{\rm loc}(z)=1-|\Phi_\sigma(bz)|^2
=b^2z^2\operatorname{Var}_0[\sigma]+O(z^4).
\label{eq:experimental_variance_slope}
\end{equation}
The measured quantities differ in how they grow with distance.  The texture $\mathcal C$ rises linearly in $z$.  The variance $\mathcal V=\mathcal C^2+\langle s_3^2\rangle_z$ rises quadratically and, at leading order, carries the same information as $\mathcal C^2$; its excess $\langle s_3^2\rangle_z$ begins at order $z^4$.  Local circular polarization is therefore better measured through $s_3$ itself, which for the real axisymmetric inputs of Sec.~\ref{sec:graded} changes sign across the beam, grows as $z^2$, and leaves the integrated $S_3$ unchanged.  The practical advantage of $\mathcal V$ over $\mathcal C$ is that it is insensitive to an uncalibrated uniform retardance of the exit optics.

\emph{Reference measurements.} A meaningful test needs at least a reference region without sample, a calibrated region of uniform twist, and a graded region with the same beam-weighted mean $\langle\Omega\rangle_0$.  With equal means, the leading average polarization angles agree while the variances differ.  With a separable input, the uniform region gives $\mathcal C=\mathcal V=0$ at leading paraxial order even while its Gaussian width or $\pm\ell$ petal pattern changes; measured with the same optics, it also serves as the null reference for the noise bias.  The graded region gives $\mathcal C\ne0$ and $\mathcal V\ne0$; its signed $s_3(\rho)$ profile can be compared with Eq.~\eqref{eq:S3_sat}, while $\int S_3\,d^2r_\perp$ stays equal to its incident value if the two circular sectors propagate without loss or conversion.  Measurements at several propagation lengths (or calibrated effective lengths) test how the signals grow, without relying on a single final image. For axisymmetric ($m=0$) linearly polarized inputs, reversing the signed defect profile exchanges the roles of the two circular sectors exactly within the leading-paraxial theory: the local polarization rotation relative to the input axis and $s_3$ change sign, while $\mathcal C$ and $\mathcal V$ are unchanged. This comparison requires the same sample interfaces and analyzer calibration.  Together, these observations test the model, but polarimetry alone cannot identify microscopic screw dislocations.

\emph{Beam waist and position.} Changing the waist or the transverse position of the beam changes the sampled distribution, which gives a quantitative test of the variance law.  Equation~\eqref{eq:C_sigma_general} holds for any spatial envelope with uniform linear input polarization (Appendix~\ref{app:graded_short}), so the onset slope per unit physical length is $d\mathcal C/dz|_{0^+}=2\sqrt{\operatorname{Var}_0[\Omega]}$.  For a Gaussian $I_0\propto\exp(-2|\mathbf r_\perp-\mathbf d|^2/w_0^2)$ centered at distance $d$ from the axis of the saturating profile, Gaussian integration gives the closed form
\begin{equation}
\begin{aligned}
\frac{1}{|\Delta\Omega|}\left.\frac{d\mathcal C}{dz}\right|_{0^+}
&=2\left(M_2-M_1^2\right)^{1/2},\\
M_\beta&=\frac{\alpha}{\alpha+\beta}
\exp\!\left(-\frac{\alpha\beta}{\alpha+\beta}\frac{d^2}{R^2}\right),
\qquad \alpha=\frac{2R^2}{w_0^2},
\end{aligned}
\label{eq:waist_slope}
\end{equation}
which for $d=0$ reduces to Eq.~\eqref{eq:C_sat_slope} written per unit length.  On axis, the slope vanishes both for $w_0\ll R$, where the beam samples only the nearly uniform core, and for $w_0\gg R$, where it samples mainly the asymptotic plateau.  It is maximal at $s^2=(\sqrt5-1)/4$, i.e., at $w_0\simeq1.80R$, where it is about twice its value at the reference waist used below [Fig.~\ref{fig:experiment_readout}(a)].  For beams narrower than about $1.6R$, displacing the beam toward the steepest part of the profile, $d\simeq R/\sqrt2$, increases the slope further; at the reference waist the gain is a factor of $1.49$.  Full radial and two-dimensional propagation of the off-axis beam, including its magnetic coupling, reproduces Eq.~\eqref{eq:waist_slope} to about $10^{-4}$ relative error.  Since the uniform control gives zero for every waist and position, the measured dependence on the waist characterizes the graded profile.

\emph{Numerical example.} For the saturating profile of Eq.~\eqref{eq:sigma_sat}, take a vacuum wavelength $\lambda_0=1.064\,\mu{\rm m}$, an amplitude waist $w_0=10\,\mu{\rm m}$, and the dimensionless parameters $(\Gamma_0,\Delta\Gamma,s)=(0.7,2.25,1.5)$ used above. These correspond to $k_0=5.91\times10^6\,{\rm m}^{-1}$, $Q=k_0w_0\simeq59.1$, $z_D\simeq0.591\,{\rm mm}$, a grading radius $R=15\,\mu{\rm m}$, and geometric twist scales $\Omega_0\simeq1.19\times10^3\,{\rm m}^{-1}$ and $\Delta\Omega\simeq3.81\times10^3\,{\rm m}^{-1}$. For a Gaussian $I_0\propto e^{-2r^2}$, its beam-weighted mean is
\begin{equation}
\langle V_s\rangle_0
=\Gamma_0+\frac{\Delta\Gamma}{2s^2+1}
=1.10909\ldots,
\label{eq:experimental_meancontrol}
\end{equation}
which specifies the uniform control with the same initial mean rotation rate. At $\zeta=0.08$ the model propagation length is $L\simeq47.2\,\mu{\rm m}$ and the common leading adapted-frame angle is $-\langle V_s\rangle_0\zeta\simeq-0.0887$ rad. The graded case instead has initial texture slope $d\mathcal C/d\zeta|_{0^+}=0.680768\ldots$: the leading short-distance prediction is $\mathcal C\simeq0.0545$, and numerical propagation of Eq.~\eqref{eq:sat_radial} gives $\mathcal C\simeq0.0535$ at this distance. The same calculation yields $\mathcal V\simeq2.94\times10^{-3}$, of which the circular excess $\langle s_3^2\rangle_z\simeq8.2\times10^{-5}$ is only $2.8\%$, and a local central circularity $s_3(0)\simeq-0.0124$, compared with the $O(\zeta^2)$ prediction $-0.0128$; the full radial $s_3$ profile at this distance is the one shown in Fig.~\ref{fig:sat_s3}. The uniform control with the same mean produces no $\mathcal C$, $\mathcal V$, or $s_3$.  Figure~\ref{fig:experiment_readout}(b) shows that each quantity follows its leading power law over this range. With the same $R$ and $\Delta\Omega$, the optimal on-axis waist $w_0\simeq27\,\mu$m doubles the texture at the same length: full propagation gives $\mathcal C\simeq0.107$ at $L\simeq47\,\mu$m. For orientation only, $\Omega=b\sigma/2$ with a Burgers-vector magnitude $b\simeq0.5$\,nm corresponds to $\sigma_0\simeq4.7\times10^{8}\,{\rm cm}^{-2}$ and $\Delta\sigma\simeq1.5\times10^{9}\,{\rm cm}^{-2}$; this conversion characterizes the metric, not the optical response of a crystal with that dislocation density. These numbers are parameters for an effective medium or an emulator, not predictions for an existing sample.

Agreement of $\mathcal C$, $\mathcal V$, and the local $s_3$ with these predictions would confirm the effective propagation equation.  Showing that a sample realizes the metric Maxwell model would also require independent measurements of its constitutive response and interfaces, and inferring an actual dislocation density would require a model that relates the microstructure of the specific material to its optical response.

\section{Conclusions}

We have rederived the propagation of structured light in screw-dislocation geometries from minimally coupled Maxwell theory, keeping $F=dA$ throughout.  The main result is the different role of a uniform density and of density gradients.  A uniform screw-dislocation density has a constant orthonormal defect strength $f'/\rho=b\sigma_0$ and produces no radius-dependent local helicity coupling; the polarization of an axial beam rotates uniformly, and finite beams undergo Landau-type orbital dynamics.

For a uniform density, we obtained the closed transverse propagator for an arbitrary input and estimated its accuracy against exact Maxwell modes, separately for the propagation shift, the longitudinal field, and the transverse profile; all three errors vanish with the paraxial parameter.  The leading paraxial evolution factorizes into an orbital unitary and a global polarization rotation, so spatial--polarization separability is preserved.  Two observables depend on the geometry alone: the breathing of a Gaussian beam, whose width is constant for the matched waist, and the rotation of OAM petal patterns.  Balanced $\pm\ell$ superpositions rotate at $-\Omega_g$ for every $\ell$, together with the uniform polarization axis, without spin--orbit conversion.

A nonuniform density changes the picture.  For an arbitrary smooth axisymmetric $\sigma(\rho)$, the polarization coherence in the local-phase limit is the characteristic function $\langle e^{-ibz\sigma}\rangle$ of the density sampled by the incident beam.  The beam-weighted mean density sets the mean rotation, the density variance sets the universal initial texture slope, and higher cumulants give the later local-phase corrections.  Circular polarization produced by diffraction appears one order later and redistributes helicity locally, while the integrated circular Stokes component is conserved in the lossless leading-paraxial theory.  A saturating density gives a closed hypergeometric expression, and numerical results for other bounded profiles confirm that the onset law does not depend on the profile.  The leading Hamiltonian also has a Pauli form with regular modes of zero propagation shift, and for coherent inputs the linear entropy of the reduced polarization state is exactly related to the measured linear-Stokes texture and local circular polarization.  A radial calculation with the full graded Maxwell equations shows that the low-lying forward-mode shifts converge to the paraxial spectrum as $Q$ increases; it does not establish a general nonparaxial texture theorem.

Calibrated full-Stokes imaging can test these predictions.  The full-Stokes variance equals twice the linear entropy of the reduced polarization state for a coherent input and is insensitive to uniform retardance after the sample, and a comparison of uniform and graded samples with the same mean twist separates a common polarization rotation from texture produced by the density variance.  The dependence of the onset slope on the beam waist and offset follows in closed form; for the saturating profile it is largest on axis near $w_0\simeq1.8R$, and it vanishes for the uniform control.  A numerical example gives the expected texture and local circular polarization, together with the requirements on the constitutive response and the interfaces.  In the leading paraxial metric-Maxwell theory, a uniform defect geometry thus shows up through geometric rotation and Landau orbital dynamics, whereas polarization textures indicate gradients of the defect density or additional constitutive structure.  The earlier applications of the same metric need different corrections: the optical laws of Ref.~\cite{SilvaPRA2026} rest on the coordinate and connection problems discussed in Sec.~\ref{sec:relation_prior}, whereas the finite-dot result of Ref.~\cite{SilvaOQE2026} has a boundary problem and leaves the translational Landau structure intact.  Applying these results to a specific natural material requires a separately justified constitutive model.  Natural extensions are finite interfaces, nonaxisymmetric gradients, and material-specific constitutive terms; they can be added without changing the distinction established here between uniform geometric transport and gradient-induced polarization structure.

\section*{Data availability}
All results in this article can be reproduced from the equations, parameters, and numerical methods given in the text and in Appendix~\ref{app:numerics}.  No experimental data were used.  The numerical data and scripts used for the figures and tables are available upon reasonable request from the author.

\begin{acknowledgments}
This work was supported by Conselho Nacional de Desenvolvimento Cient\'ifico e Tecnol\'ogico (CNPq) (grant 305427/2026-1), Funda\c c\~ao de Amparo \`a Pesquisa e ao Desenvolvimento Cient\'ifico e Tecnol\'ogico do Maranh\~ao (FAPEMA) (grant UNIVERSAL-06395/22), and Coordena\c c\~ao de Aperfei\c coamento de Pessoal de N\'ivel Superior -- Brasil (CAPES) -- Finance Code 001.
\end{acknowledgments}

\appendix

\section{Metric identities}
\label{app:geometry}

The metric block in the $(\phi,z)$ sector of Eq.~\eqref{eq:metric} is
\begin{equation}
\begin{pmatrix}\rho^2+f^2&f\\ f&1\end{pmatrix},
\end{equation}
whose determinant is $\rho^2$.  Direct inversion gives Eq.~\eqref{eq:inverse_det}.  The dual frame follows from $\vartheta^{\hat a}(e_{\hat b})=\delta^{\hat a}{}_{\hat b}$,
\begin{equation}
e_{\hat1}=\partial_\rho,\qquad
e_{\hat2}=\rho^{-1}\partial_\phi-\frac{f}{\rho}\partial_z,\qquad
e_{\hat3}=\partial_z.
\end{equation}
The exterior derivatives are
\begin{equation}
d\vartheta^{\hat2}=\rho^{-1}\vartheta^{\hat1}\wedge\vartheta^{\hat2},\qquad
d\vartheta^{\hat3}=\frac{f'}{\rho}\vartheta^{\hat1}\wedge\vartheta^{\hat2}.
\end{equation}
For a transverse annulus between the axis/core and a circle of radius $\rho$, Stokes' theorem gives $\int d\vartheta^{\hat3}=\oint_{\partial D_\rho}\vartheta^{\hat3}-\lim_{\epsilon\to0}\oint_{\partial D_\epsilon}\vartheta^{\hat3}=2\pi[f(\rho)-f_0]$.  Equating this to the signed continuum Burgers flux $bN(\rho)$ proves Eqs.~\eqref{eq:Burgers_flux} and \eqref{eq:defect_map}; in the regular continuum sector used in the main text, $f_0=0$.  This also makes the cylindrical scale factor explicit: $d\vartheta^{\hat3}=b\sigma\,\vartheta^{\hat1}\wedge\vartheta^{\hat2}=b\sigma\rho\,d\rho\wedge d\phi$.
Substituting these expressions into the torsion-free Cartan equation gives the Levi--Civita connection quoted in Sec.~\ref{sec:geometry}.  Subtracting it from the auxiliary defect connection $\widetilde\omega^{\hat1}{}_{\hat2}=-\rho^{-1}\vartheta^{\hat2}$ gives the contortion one-forms listed there and reproduces Eq.~\eqref{eq:torsion}.

For uniform $f=\Omega_g\rho^2$, use $\rho^2d\phi=x\,dy-y\,dx$.  The spatial metric is invariant under
\begin{equation}
\xi_1=\partial_x-\Omega_g y\partial_z,\qquad
\xi_2=\partial_y+\Omega_g x\partial_z,\qquad
\xi_3=\partial_z,
\end{equation}
as direct substitution gives $\mathcal L_{\xi_i}g=0$.  These three independent Killing fields act transitively, which proves the homogeneity used in the main text.

For completeness, if $\widetilde\Gamma^\alpha{}_{\mu\nu}=\Gamma^\alpha{}_{\mu\nu}+K^\alpha{}_{\mu\nu}$, then for an antisymmetric tensor
\begin{equation}
\widetilde\nabla_\mu F^{\mu\nu}-\nabla_\mu F^{\mu\nu}
=K^\mu{}_{\mu\lambda}F^{\lambda\nu}+K^\nu{}_{\mu\lambda}F^{\mu\lambda}.
\end{equation}
The torsion trace vanishes for the chosen defect connection, while antisymmetry of $F$ gives $K^\nu{}_{\mu\lambda}F^{\mu\lambda}=\tfrac12T^\nu{}_{\mu\lambda}F^{\mu\lambda}$.  With $T^z{}_{\rho\phi}=f'$, Eq.~\eqref{eq:divdifference} follows.  This identity is used only for comparison; the electromagnetic equations remain Eq.~\eqref{eq:maxwell_cov}.

For the wave equation, write $F^{\mu\nu}=\nabla^\mu A^\nu-\nabla^\nu A^\mu$ in the source-free Maxwell equation.  Commuting the derivatives with the curvature convention stated below Eq.~\eqref{eq:maxwell_cov} and imposing $\nabla_\mu A^\mu=0$ gives Eq.~\eqref{eq:wave}.  A WKB substitution $A^\mu=(a^\mu+\epsilon b^\mu+\cdots)e^{iS/\epsilon}$ then yields the $\epsilon^{-2}$ equations in Eq.~\eqref{eq:eikonal} and, at order $\epsilon^{-1}$, the transport equation quoted before Eq.~\eqref{eq:transport}.  Evaluating the Levi--Civita connection on the axial null tangent $e_{\hat0}+e_{\hat3}$ gives the transverse matrix in Sec.~\ref{sec:wkb} and hence Eq.~\eqref{eq:rotation}.

\section{Curl and divergence}
\label{app:curl}

For $U^\flat=U_\rho\vartheta^{\hat1}+U_\phi\vartheta^{\hat2}+U_z\vartheta^{\hat3}$ with $e^{i(m\phi+k_zz)}$ dependence, direct expansion gives
\begin{align}
dU^\flat={}&(\Cf\mathbf U)_z\,\vartheta^{\hat1}\wedge\vartheta^{\hat2}
-(\Cf\mathbf U)_\phi\,\vartheta^{\hat1}\wedge\vartheta^{\hat3}\nonumber\\
&+(\Cf\mathbf U)_\rho\,\vartheta^{\hat2}\wedge\vartheta^{\hat3},
\end{align}
with the coefficients in Eq.~\eqref{eq:curl}.  Applying the spatial Hodge operator and $d$ once more gives Eq.~\eqref{eq:div} and the identity $\Df\Cf=0$ because $d^2=0$.  With the $e^{-i\omega t}$ convention, the exterior Maxwell equations then give Eq.~\eqref{eq:maxwell_harmonic}.

\section{Derivation of the paraxial Hamiltonian}
\label{app:paraxial}

A regular Cartesian form of the transverse coframe is useful for deriving Eq.~\eqref{eq:paraxial} without passing through a scalar wave equation.  Write
\begin{align}
\vartheta^{\hat3}&=dz+A_xdx+A_ydy,\nonumber\\
\partial_xA_y-\partial_yA_x&=B_g=2\Omega(x,y).
\end{align}
For a longitudinal Fourier factor $e^{ik_zz}$ the horizontal derivatives are
\begin{equation}
D_i=\partial_i-ik_zA_i,
\qquad
[D_x,D_y]=-ik_zB_g.
\end{equation}
The exact curl of a physical vector $\mathbf U$ is
\begin{align}
(\Cf\mathbf U)_x&=D_yU_z-ik_zU_y,\nonumber\\
(\Cf\mathbf U)_y&=ik_zU_x-D_xU_z,\nonumber\\
(\Cf\mathbf U)_z&=D_xU_y-D_yU_x+B_gU_z.
\label{eq:cartcurl_app}
\end{align}
Gauss's law gives
\begin{equation}
D_xE_x+D_yE_y+ik_zE_z=0,
\qquad
E_z=\frac{i}{k_z}(D_xE_x+D_yE_y).
\label{eq:Ez_app}
\end{equation}
Applying the curl twice and using Eq.~\eqref{eq:Ez_app}, the exact transverse part of $\Cf^2\mathbf E=k_0^2\mathbf E$ becomes
\begin{align}
(\Cf^2\mathbf E)_\perp={}&
\left(k_z^2-D_x^2-D_y^2\right)\mathbf E_\perp
-k_zB_g\sigma_y\mathbf E_\perp\nonumber\\
&+\begin{pmatrix}
B_gD_yE_z+(\partial_yB_g)E_z\\
-B_gD_xE_z-(\partial_xB_g)E_z
\end{pmatrix}.
\label{eq:C2trans_app}
\end{align}
The Pauli matrix $\sigma_y$ is diagonal in the circular basis of Eq.~\eqref{eq:circular}.  For a forward beam, set $k_z=k_0-\beta$, replace $k_z$ by $k_0$ inside the already subleading transverse covariant derivatives, and use $k_z^2-k_0^2=-2k_0\beta+O(\beta^2)$.  Neglecting the last line of Eq.~\eqref{eq:C2trans_app} then gives
\begin{equation}
\beta\Psi=
\left[-\frac{1}{2k_0}(\nabla_\perp-ik_0\Ag)^2\I
-\frac{B_g}{2}\sigma_3\right]\Psi,
\end{equation}
which is Eq.~\eqref{eq:paraxial} because $B_g=2\Omega$.

The omitted terms are parametrically smaller.  With a beam scale $w$, Eq.~\eqref{eq:Ez_app} gives $E_z/E_\perp=O[(k_0w)^{-1}]$.  Compared with the transverse kinetic term $D_\perp^2E_\perp\sim E_\perp/w^2$, the $B_gD_\perp E_z$ terms are of relative order $|\Omega|/k_0$.  If the geometry varies on a transverse scale $L_g$, the $(\nabla B_g)E_z$ terms are of relative order $(|\Omega|/k_0)(w/L_g)$.  At fixed $\Gamma=\Omega k_0w^2$ both scale as $Q^{-2}$, with $Q=k_0w$; away from zeros of $\Omega$ they are also smaller than the helicity term $k_0B_gE_\perp$ by $O(Q^{-2})$ and $O[Q^{-2}(w/L_g)]$.  The replacement $k_z\to k_0$ inside already subleading terms and the discarded $\beta^2$ contribution add corrections controlled by $|\Omega|/k_0$ and $Q^{-2}$.  These estimates give Eq.~\eqref{eq:graded_vector_error} and justify the slow-gradient condition used in Sec.~\ref{sec:graded}.

\section{Pauli identity and zero-shift Maxwell modes}
\label{app:pauli_modes}

For $\Pi_\pm=\Pi_x\pm i\Pi_y$, the two operator products are
\begin{equation}
\Pi_-\Pi_+=\Pi_x^2+\Pi_y^2-k_0B_g,\qquad
\Pi_+\Pi_-=\Pi_x^2+\Pi_y^2+k_0B_g.
\end{equation}
Consequently, $H_+=\Pi_-\Pi_+/(2k_0)$ and $H_-=\Pi_+\Pi_-/(2k_0)$ as in Eq.~\eqref{eq:pauli_factorization}.  In polar coordinates, using $A_{g,\phi}=f(\rho)/\rho$,
\begin{equation}
\Pi_+=-ie^{i\phi}\!\left(\partial_\rho+\frac{i}{\rho}\partial_\phi+\frac{k_0f(\rho)}{\rho}\right).
\end{equation}
For $u=e^{im\phi}R_m(\rho)$ the equation $\Pi_+u=0$ reduces to
\begin{equation}
R_m'=\left(\frac{m}{\rho}-\frac{k_0 f(\rho)}{\rho}\right)R_m,
\end{equation}
whose regular solution is Eq.~\eqref{eq:pauli_zero_mode}.  On the infinite transverse plane with positive asymptotic density $\Omega_\infty>0$, $f\sim\Omega_\infty\rho^2$ and the amplitude decays as $\rho^m e^{-k_0\Omega_\infty\rho^2/2}$.  If the metric field has finite total flux instead, square integrability must be decided from the full asymptotic of $f$.  These statements assume the regular axis, exclude singular-core boundary conditions, and do not establish zero modes for a finite hard-wall aperture.

These modes have a counterpart in the first-order Beltrami system.  Put $E_z=0$, $k_z=k_0$, and select the positive-helicity transverse component.  Then its transverse equations imply $E_\phi=iE_\rho$, while the longitudinal equation is $\Pi_+\psi_+=0$.  Therefore, the normalizable $\psi_+$ family of Eq.~\eqref{eq:pauli_zero_mode} also solves the source-free, untruncated Maxwell equations in this exceptional sector.  This holds for these particular modes only; general paraxial modes are not exact Maxwell solutions.

\section{Derivation of the uniform Landau propagator}
\label{app:propagator}

Equation~\eqref{eq:Hosc} separates because $[H_{\rm osc},L_z]=0$,
\begin{equation}
e^{-izH_{\rm orb}}=e^{-izH_{\rm osc}}e^{+i\Omega_gzL_z}.
\end{equation}
The isotropic two-dimensional oscillator kernel is
\begin{align}
K_{\rm osc}(\mathbf r,\mathbf s;z)
={}&\frac{k_0\Omega_g}{2\pi i\sin(\Omega_g z)}\nonumber\\
&\times\exp\Bigg\{\frac{i k_0\Omega_g}{2\sin(\Omega_g z)}
\Big[(r^2+s^2)\cos(\Omega_g z)\nonumber\\
&\hspace{39mm}-2\mathbf r\cdot\mathbf s\Big]\Bigg\}.
\end{align}
The second factor rotates the input argument, so the orbital kernel is $K_{\rm osc}(\mathbf r,R_{-\Omega_gz}\mathbf r';z)$.  Using
\begin{equation}
\mathbf r\cdot R_{-\alpha}\mathbf r'
=\cos\alpha\,\mathbf r\cdot\mathbf r'
+\sin\alpha\,(\mathbf r\times\mathbf r')_z
\end{equation}
gives Eq.~\eqref{eq:Landau_kernel}.  Its $\Omega_g\rightarrow0$ limit is the ordinary Fresnel kernel $k_0[2\pi iz]^{-1}\exp[ik_0|\mathbf r-\mathbf r'|^2/(2z)]$.  Direct substitution verifies $i\partial_z\mathcal K_\Omega=H_{\rm orb}\mathcal K_\Omega$ identically away from the caustic planes.  At $\Omega_g z=n\pi$ the Mehler representation is understood distributionally; the evolution remains unitary because it is the product of the oscillator and rotation unitaries.

\section{Uniform orbital spectrum and Gaussian width}
\label{app:uniform_orbital}

For $H_{\rm osc}$ in Eq.~\eqref{eq:Hosc}, separation $u=R(\rho)e^{im\phi}$ gives
\begin{equation}
-\frac{1}{2k_0}\left(R''+\frac{R'}{\rho}-\frac{m^2}{\rho^2}R\right)
+\frac{k_0\Omega_g^2\rho^2}{2}R=E_{\rm osc}R.
\end{equation}
With $x=k_0|\Omega_g|\rho^2$ and $R=x^{|m|/2}e^{-x/2}L_n^{(|m|)}(x)$, the associated-Laguerre equation gives
\begin{equation}
E_{\rm osc}=|\Omega_g|(2n+|m|+1).
\end{equation}
Since $L_z e^{im\phi}=m e^{im\phi}$, Eq.~\eqref{eq:orb_eigs_general} follows immediately.

The Gaussian width can be obtained without an ansatz for the propagated field.  In the dimensionless variables of Sec.~\ref{sec:paraxial}, one Cartesian oscillator obeys Eq.~\eqref{eq:heisenberg_width}.  For $u(x,0)\propto e^{-x^2}$, the three moments stated below that equation give
\begin{equation}
4\langle x^2(\zeta)\rangle=
\cos^2(\Gamma\zeta)+\frac{4}{\Gamma^2}\sin^2(\Gamma\zeta).
\end{equation}
Rotational symmetry gives the same width in $y$, proving Eq.~\eqref{eq:width}.  The condition for the coefficient of $\sin^2(\Gamma\zeta)$ to equal unity is $|\Gamma|=2$, which proves Eq.~\eqref{eq:stationary}.

\section{Exact radial equation and spectrum}
\label{app:exact_modes}

From Eq.~\eqref{eq:maxwell_harmonic}, $\Cf^2\mathbf E=k_0^2\mathbf E$.  In a Beltrami sector $\Cf\mathbf E_\eta=\eta k_0\mathbf E_\eta$, let $u_\eta=E_{\eta,z}$ and $D=k_0^2-k_z^2$.  The first two components of Eq.~\eqref{eq:curl} form a $2\times2$ linear system whose inversion gives
\begin{align}
E_{\eta,\rho}&=\frac{i[k_z u_\eta'+\eta k_0 q u_\eta]}{D},\nonumber\\
E_{\eta,\phi}&=-\frac{\eta k_0 u_\eta'+k_zq u_\eta}{D}.
\label{eq:transverse_reconstruct_app}
\end{align}
Substituting Eq.~\eqref{eq:transverse_reconstruct_app} into the longitudinal Beltrami component and using $q'= -m/\rho^2-\Omega_gk_z$ for a uniform defect density gives Eq.~\eqref{eq:radial_exact}.  For $a=\Omega_gk_z>0$, set $x=a\rho^2$ and
\begin{equation}
u_\eta=x^{|m|/2}e^{-x/2}F_\eta(x).
\end{equation}
The radial equation becomes Kummer's equation~\cite{NISTHandbook}.  Suppression of the $e^{+x/2}$ asymptotic term requires $F_\eta=L_n^{(|m|)}(x)$ with $n=0,1,2,\ldots$, and the remaining constant term gives Eq.~\eqref{eq:exact_spectrum}.  Solving its quadratic equation for the forward root and expanding in $|\Omega_g|/k_0$ gives Eq.~\eqref{eq:exact_expand}.

\section{Longitudinal and transverse Landau indices}
\label{app:ladder}

The first-order Maxwell reconstruction also explains the index shift in Eq.~\eqref{eq:Nladder}.  For a Beltrami eigenfield $\Cf\mathbf E_\eta=\eta k_0\mathbf E_\eta$ with longitudinal component $u_\eta(\rho)e^{im\phi}$, the transverse components away from $D=k_0^2-k_z^2=0$ are
\begin{align}
E_{\eta,\rho}&=\frac{i[k_z u_\eta'+\eta k_0 q u_\eta]}{D},\nonumber\\
E_{\eta,\phi}&=-\frac{\eta k_0 u_\eta'+k_zq u_\eta}{D}.
\end{align}
Using Eq.~\eqref{eq:circular}, the dominant forward circular combinations are proportional to
\begin{align}
\psi_+&\propto e^{i(m-1)\phi}(u_+'+q u_+),\qquad \eta=+1,\nonumber\\
\psi_-&\propto e^{i(m+1)\phi}(u_-'-q u_-),\qquad \eta=-1.
\end{align}
For $\Omega_gk_z>0$, write $x=\Omega_gk_z\rho^2$ and $u_\eta\propto x^{|m|/2}e^{-x/2}L_n^{(|m|)}(x)$.  Standard Laguerre derivative and recurrence identities~\cite{NISTHandbook} show that the first operator raises the transverse Landau index by two, while the second lowers it by two,
\begin{equation}
N_\perp=N_{nm}+2\eta.
\end{equation}
This result holds for either sign of $m$; when $\eta=-1$ acts on the lowest $N_{nm}=1$ longitudinal rung, the lowering combination vanishes.  This is the origin of the exceptional case discussed in Appendix~\ref{app:exceptional}.

\section{Exact vector field of the lowest mode}
\label{app:maxwell_paraxial_benchmark}

Consider the nonexceptional uniform branch $\eta=+1$, $n=m=0$, for which $N=1$.  Equation~\eqref{eq:exact_spectrum} gives, for $0<\gamma<1/2$,
\begin{equation}
k_z=k_0-2\Omega_g,
\qquad
a\equiv\Omega_gk_z>0,
\end{equation}
and the longitudinal field is
\begin{equation}
u(\rho)=C e^{-a\rho^2/2},\qquad q=-a\rho.
\end{equation}
Substitution into the exact reconstruction formula~\eqref{eq:transverse_reconstruct_app}, with $D=k_0^2-k_z^2$, gives
\begin{equation}
E_\rho=-\frac{i k_z\rho}{2}u,
\qquad
E_\phi=\frac{k_z\rho}{2}u.
\label{eq:fundamental_reconstruct}
\end{equation}
No paraxial approximation has been used in Eq.~\eqref{eq:fundamental_reconstruct}.  Since
\begin{equation}
\int_0^\infty \rho e^{-a\rho^2}d\rho=\frac{1}{2a},
\qquad
\int_0^\infty \rho^3 e^{-a\rho^2}d\rho=\frac{1}{2a^2},
\end{equation}
the angularly integrated electric norms satisfy
\begin{equation}
\frac{\|\mathbf E_\perp\|^2}{\|E_z\|^2}
=\frac{k_z^2}{2a}=\frac{k_z}{2\Omega_g}.
\end{equation}
Using $k_z+2\Omega_g=k_0$ immediately gives Eq.~\eqref{eq:fz_exact}.  In the Beltrami sector, $\mathbf B=-i\eta\mathbf E$ in the scaled magnetic-field convention, so the magnetic field carries the same component ratio.

The exact transverse circular profile in Eq.~\eqref{eq:fundamental_reconstruct} is proportional to $\rho e^{-\alpha_{\rm ex}\rho^2/2}e^{-i\phi}$ with $\alpha_{\rm ex}=\Omega_gk_z$.  The corresponding normalized $\ell=-1$, $p=0$ Landau profile ($N_\perp=3$) at a positive scale $\alpha$ can be written as
\begin{equation}
\varphi_\alpha(\rho,\phi)=\frac{\alpha}{\sqrt\pi}\,
\rho e^{-\alpha\rho^2/2}e^{-i\phi}.
\end{equation}
Hence
\begin{equation}
\langle\varphi_\alpha|\varphi_\beta\rangle
=\frac{4\alpha\beta}{(\alpha+\beta)^2}.
\end{equation}
Taking $\alpha=\Omega_gk_0$ and $\beta=\Omega_gk_z=\kappa\alpha$ and squaring the overlap yields Eq.~\eqref{eq:profile_fidelity}.  This mode thus tests three separate paraxial approximations: the propagation shift, the neglected longitudinal component, and the transverse radial scale.

\section{Exceptional Maxwell sector}
\label{app:exceptional}

For a Beltrami sector with eigenvalue $\lambda$ and $D=\lambda^2-k_z^2=0$, write $\lambda=s k_z$, $s=\pm1$.  The first two Beltrami equations give
\begin{equation}
u'+squ=0,
\end{equation}
so for uniform positive $\Omega_gk_z$ the decaying longitudinal solution occurs for $s=-1$,
\begin{equation}
u=C\rho^m e^{-\Omega_gk_z\rho^2/2},\qquad m\ge0.
\end{equation}
The remaining transverse equation is
\begin{equation}
E_\rho'+\left(\frac1\rho-sq\right)E_\rho
=-i\left(k_z+\frac{q^2}{k_z}\right)u.
\label{eq:exceptional_transverse}
\end{equation}
For the decaying $s=-1$ solution, the integrating factor is $\mu(\rho)=\rho^{m+1}e^{-a\rho^2/2}$ with $a=\Omega_gk_z>0$.  Regularity at the origin fixes the homogeneous integration constant.  Normalizability at infinity would then require the total source integral to vanish, but
\begin{align}
J&=\int_0^\infty d\rho\,\rho^{2m+1}e^{-a\rho^2}
\left[k_z+\frac{(m/\rho-a\rho)^2}{k_z}\right]\nonumber\\
&=\frac{\Gamma(m+1)}{2a^{m+1}}
\left(k_z+\frac{a}{k_z}\right)>0.
\label{eq:exceptional_integral}
\end{align}
The solution regular at the axis therefore has an exponentially growing transverse tail, and this longitudinal solution is not a physical guided Maxwell mode.

A physical transverse exceptional family is obtained instead by setting $u=E_z=0$ in the first-order Beltrami equations.  For $k_z=k_0$, the forward branch, and the circular convention used in the main text,
\begin{equation}
\psi_+\propto \rho^\ell e^{i\ell\phi}e^{-\Omega_gk_0\rho^2/2},
\qquad \psi_-=0,\qquad \ell\ge0.
\end{equation}
Its transverse energy is finite.  This family coincides with the zero-shift Landau family of Eq.~\eqref{eq:Huniform}, which checks the paraxial reduction at the vector level.

\section{Short-distance expansion for an arbitrary density}
\label{app:graded_short}

Write the dimensionless circular-sector Hamiltonians as $H_h=H_0-hV$ with $h=\pm1$, where $H_0$ is the common orbital operator and $V=z_D\Omega$.  For equal initial circular components $\psi_h(0)=u_0/\sqrt2$,
\begin{equation}
\psi_h(\zeta)=\frac{1}{\sqrt2}
\left[u_0-i\zeta H_0u_0+i h\zeta Vu_0+O(\zeta^2)\right].
\label{eq:short_wave_expansion}
\end{equation}
The common orbital contribution cancels from the normalized Jones ratio at first order.  Hence
\begin{equation}
s_1+i s_2=e^{2i\theta_0}e^{-2iV\zeta}+O(\zeta^2).
\label{eq:short_local_coherence_app}
\end{equation}
Using $2V\zeta=2\Omega z=b\sigma z$ gives Eq.~\eqref{eq:general_local_stokes}.  Neglecting transverse redistribution while keeping the phase in Eq.~\eqref{eq:short_local_coherence_app} gives the local-phase average
\begin{equation}
\left\langle s_1+i s_2\right\rangle_{\rm loc}
=e^{2i\theta_0}\int e^{-ibz\sigma}d\mu_0,
\end{equation}
which proves Eq.~\eqref{eq:characteristic_theorem}.

The cumulant expansion follows from the standard identity
\begin{equation}
\ln\left\langle e^{-it\sigma}\right\rangle_0
=\sum_{n\ge1}\frac{(-it)^n}{n!}\kappa_n[\sigma],
\end{equation}
with $t=bz$.  Separating the imaginary and real parts proves Eqs.~\eqref{eq:mean_angle_cumulants} and \eqref{eq:texture_cumulants}.  In particular,
\begin{equation}
|\Phi_\sigma(bz)|^2
=1-b^2z^2\operatorname{Var}_0[\sigma]+O(z^4),
\end{equation}
so the local-phase approximation gives
\begin{equation}
\mathcal C_{\rm loc}=|b|\,|z|\sqrt{\operatorname{Var}_0[\sigma]}+O(|z|^3).
\end{equation}
Because Eq.~\eqref{eq:short_wave_expansion} is the full leading-paraxial short-distance expansion, the same coefficient proves Eq.~\eqref{eq:C_sigma_general} with the conservative remainder $O(z^2)$.

Within the leading paraxial theory, conservation of the integrated $S_3$ follows without a short-distance expansion.  Each $H_h$ is Hermitian, so
\begin{equation}
\frac{d}{dz}\int d^2r\,|\psi_h|^2=0.
\end{equation}
An initially linear state has equal circular powers and so satisfies Eq.~\eqref{eq:global_S3_zero} at every propagation distance.

For the local $S_3$ onset, take a real axisymmetric input with no initial radial current.  Since $\Ag$ is azimuthal while $V$ is radial, the current term proportional to $\Ag$ has no radial divergence at $\zeta=0$.  The opposite first-order phases generate
\begin{equation}
\mathbf j_h=h\zeta\,\frac{I_0}{2}\nabla_rV+O(\zeta^2).
\end{equation}
The continuity equations $\partial_\zeta|\psi_h|^2=-\nabla_r\cdot\mathbf j_h$ give
\begin{equation}
\frac{S_3}{S_0}=-\frac{\zeta^2}{2I_0}
\nabla_r\cdot(I_0\nabla_rV)+O(\zeta^4).
\end{equation}
Using $V=z_D\Omega$, $\zeta=z/z_D$, $\nabla_r=w_0\nabla_\perp$, and $z_D=k_0w_0^2$ gives Eq.~\eqref{eq:S3_sigma_general}.  The $O(z^4)$ remainder follows because the radial Hamiltonians and the chosen real input make each component intensity even in $z$.

For the Gaussian input $I_0\propto e^{-2r^2}$ and the saturating example, $X=e^{-r^2/s^2}$ obeys
\begin{equation}
\langle X^n\rangle_0=\frac{2s^2}{2s^2+n}.
\end{equation}
Equations~\eqref{eq:var_sat} and \eqref{eq:C_sat_slope} follow immediately.  The exact local-phase integral is
\begin{align}
\frac{\int_0^\infty r\,dr\,e^{-2r^2}e^{-2iV_s(r)\zeta}}
{\int_0^\infty r\,dr\,e^{-2r^2}}
={}&2s^2 e^{-2i(\Gamma_0+\Delta\Gamma)\zeta}\nonumber\\
&\times\int_0^1du\,u^{2s^2-1}e^{2i\Delta\Gamma\zeta u}\nonumber\\
={}&e^{-2i(\Gamma_0+\Delta\Gamma)\zeta}\nonumber\\
&\times{}_1F_1(2s^2;2s^2+1;2i\Delta\Gamma\zeta),
\end{align}
which proves Eqs.~\eqref{eq:sat_localphase_average} and \eqref{eq:C_sat_exactlocal}.  Finally,
\begin{equation}
\frac{1}{I_0}\nabla_r\cdot(I_0\nabla_rV_s)
=\frac{4\Delta\Gamma}{s^2}e^{-r^2/s^2}
\left[1-\left(2+\frac1{s^2}\right)r^2\right],
\end{equation}
which proves Eq.~\eqref{eq:S3_sat}.

\section{Exact Maxwell modes for the saturating profile}
\label{app:graded_maxwell}

The derivation of the longitudinal Beltrami equation in Appendix~\ref{app:exact_modes} requires no constant-density approximation.  For a general smooth axisymmetric $f$ and $\Omega=f'/(2\rho)$, the longitudinal component $u_\eta=E_{\eta,z}$ obeys
\begin{equation}
\begin{aligned}
u_\eta''+\frac{1}{\rho}u_\eta'&-\Bigg[
\left(\frac{m-k_z f(\rho)}{\rho}\right)^2+k_z^2-k_0^2\\
&\qquad +2\eta k_0\Omega(\rho)\Bigg]u_\eta=0.
\end{aligned}
\label{eq:graded_maxwell_radial}
\end{equation}
The $2\eta k_0\Omega$ term comes from $q'+q/\rho=-k_z f'/\rho$ in the exact first-order Maxwell reconstruction.  No $f''$ term or phenomenological helicity potential appears.  Set $r=\rho/w_0$, $Q=k_0w_0$, $V(r)=z_D\Omega$, $z_D=k_0w_0^2$, and $F(r)=Q f(\rho)/w_0=2\int_0^r sV(s)\,ds$.  With longitudinal momentum in units of $w_0^{-1}$, $K=k_zw_0=Q-\beta w_0$, Eq.~\eqref{eq:graded_maxwell_radial} becomes the nonlinear spectral problem
\begin{equation}
\begin{aligned}
\Bigg[-\partial_r^2-\frac1r\partial_r
&+\frac{[m-(K/Q)F(r)]^2}{r^2}\\
&+2\eta V(r)\Bigg]u_\eta=(Q^2-K^2)u_\eta.
\end{aligned}
\label{eq:graded_maxwell_nonlinear}
\end{equation}
At leading paraxial order $K/Q\to1$, and the transverse circular index is $m'=m-\eta$ (the ladder mapping of Appendix~\ref{app:ladder}).  The corresponding paraxial scalar eigenproblem is
\begin{equation}
\begin{aligned}
\Bigg[-\frac12\left(\partial_r^2+\frac1r\partial_r\right)
&+\frac12\left(\frac{m'}r-\frac{F(r)}r\right)^2\\
&-\eta V(r)\Bigg]\psi_\eta=\varepsilon_p\psi_\eta,\\
\beta_p w_0&=\frac{\varepsilon_p}{Q}.
\end{aligned}
\label{eq:graded_paraxial_radial}
\end{equation}
Both radial operators are discretized in conservative finite-volume form on a midpoint grid, symmetrized with the radial area weight, and solved as real symmetric tridiagonal eigenproblems.  For each root $K$, the potential in Eq.~\eqref{eq:graded_maxwell_nonlinear} is updated self-consistently until $|\Delta K|/Q<10^{-13}$.  We use $r_{\max}=9$ and the smooth saturating $V$ of Eq.~\eqref{eq:Vsat} at $(\Gamma_0,\Delta\Gamma,s)=(0.7,2.25,1.5)$.  The scheme is second order in the cell size, and the discretization error of the relative shift error is nearly independent of $Q$ (about $7\times10^{-7}$ for $N=6000$ cells), so it must be removed at large $Q$.  Each relative error is computed with $N=12000$ and $N=24000$ cells and Richardson-extrapolated; the extrapolation uncertainty, estimated from the $N=6000$/$12000$ pair, is below $3\times10^{-9}$, and enlarging the domain to $r_{\max}=12$ changes the results by less than $2\times10^{-10}$.  The lowest $\eta=+$ paraxial mode for $m'\ge0$ is a physical zero-shift mode and is excluded from a relative-error ratio.  For $\eta=-$, the longitudinal exceptional lowest rung annihilates the dominant circular transverse component and is excluded; the next four regular branches are compared instead.

\begin{table}[b]
\caption{Maximum absolute relative shift error $\max|\beta_p-\beta_{\rm ex}|/|\beta_p|$ over four nonzero modes for each listed $(m,\eta)$ sector (eight in total for $\eta=+1$).  Grid-converged values (Richardson extrapolation from $N=12000$ and $24000$ cells), rounded to three digits.}
\label{tab:graded_maxwell_errors}
\begin{ruledtabular}
\begin{tabular}{ccc}
$Q$&$\eta=+1$ ($m=1,2$)&$\eta=-1$ ($m=0$)\\
\hline
50 & $1.14\times10^{-3}$ & $2.67\times10^{-3}$\\
100 & $2.84\times10^{-4}$ & $6.66\times10^{-4}$\\
200 & $7.09\times10^{-5}$ & $1.66\times10^{-4}$\\
\end{tabular}
\end{ruledtabular}
\end{table}
\begin{figure}[t]
\includegraphics[width=\columnwidth]{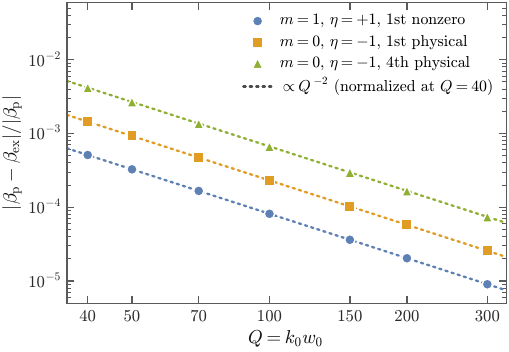}
\caption{Relative difference between the paraxial and exact Maxwell propagation shifts for three physical radial branches of the saturating profile.  Markers are grid-converged results for the labeled longitudinal index $m$, helicity $\eta$, and radial branch; dotted lines show the $Q^{-2}$ trend normalized at $Q=40$.  The positive-helicity zero-shift mode and the unphysical longitudinal exceptional branch are excluded.}
\label{fig:graded_maxwell}
\end{figure}
These values follow the $Q^{-2}$ trend but do not give a general estimate for high radial indices.  An independent calculation in a Laguerre spectral basis reproduces all entries of Table~\ref{tab:graded_maxwell_errors} to the digits shown.  Figure~\ref{fig:graded_maxwell} shows selected physical branches.  This modal check does not determine the next-order vector corrections to the density-variance law for arbitrary beams.

\section{Numerical methods}
\label{app:numerics}

Equation~\eqref{eq:sat_radial} is propagated after the radial similarity transformation $u=\sqrt r\,\psi$, which turns the radial norm into $\int|u|^2dr$ and the kinetic operator into a real symmetric tridiagonal form.  Figures~\ref{fig:sat_stokes} and \ref{fig:sat_s3} use $r_{\max}=11$ and $N=900$ cells, Fig.~\ref{fig:sat_C} uses $r_{\max}=11$ and $N=650$, Fig.~\ref{fig:general_universality} uses $r_{\max}=10$ and $N=1600$, and Fig.~\ref{fig:experiment_readout} uses $r_{\max}=9$ and $N=1600$; each circular sector is diagonalized independently and propagated by its exact spectral phase.  An independent two-dimensional split-step Fourier propagation~\cite{FeitFleck1978} reproduces the plotted texture and Stokes curves of Figs.~\ref{fig:sat_stokes}--\ref{fig:general_universality} to within $10^{-4}$.  The radial norm is conserved to machine precision.

For $(\Gamma_0,\Delta\Gamma,s)=(0.7,2.25,1.5)$, Eq.~\eqref{eq:C_sat_slope} predicts
\begin{equation}
\left.\frac{d\mathcal C}{d\zeta}\right|_{0^+}=0.6807684\ldots.
\end{equation}
The numerical ratios are $\mathcal C/\zeta=0.680764$ at $\zeta=0.002$ and $0.679978$ at $\zeta=0.02$ on the $N=1000$ grid; the spectral calculation described below gives $0.680760$ and $0.679974$.  At $\zeta=(0.4,0.8,1.2)$ the texture measure is $(0.17524,0.15468,0.13042)$ for $N=500$ and $(0.17520,0.15466,0.13045)$ for $N=1000$, so the change under a factor-of-two refinement remains below $5\times10^{-5}$.  The maximum deviation of the numerical $s_3$ from Eq.~\eqref{eq:S3_sat} in the beam core scales as $O(\zeta^4)$.

The test with several profiles in Fig.~\ref{fig:general_universality} uses a separate finite-difference implementation on $0<r<10$ with $N=1600$.  Besides the exponential saturation, the dimensionless twist profiles are
\begin{align}
V_2(r)&=0.7+2.0\frac{r^2}{1.5^2+r^2},\\
V_3(r)&=0.7+1.8\tanh\!\left(\frac{r^2}{2.0^2}\right),\\
V_4(r)&=0.7+1.5e^{-r^2/1.2^2}.
\end{align}
For each case, the metric coefficient is obtained from
\begin{equation}
a_\phi(r)=\frac{2}{r}\int_0^r r'V_j(r')\,dr',
\end{equation}
which is the dimensionless form of Eq.~\eqref{eq:general_sigma_map}.  At $\zeta=0.002$, the predicted and numerical values of $\mathcal C/\zeta$ are, respectively,
\begin{center}
\begin{tabular}{lcc}
\hline\hline
profile & prediction & numerical\\
\hline
exponential saturation & 0.680768 & 0.680761\\
Lorentz saturation & 0.491486 & 0.491482\\
$\tanh$ saturation & 0.415587 & 0.415584\\
Gaussian core & 0.593986 & 0.593980\\
\hline\hline
\end{tabular}
\end{center}
The differences between the two columns are the physical $O(\zeta^2)$ corrections to $\mathcal C/\zeta$ discussed in Sec.~\ref{sec:graded}.  As a check on the grid, the same quantities were computed in a basis of two-dimensional oscillator (Laguerre) functions, with potential matrix elements from Gauss--Laguerre quadrature~\cite{NISTHandbook} and exact spectral propagation.  This gives $0.680760$, $0.491481$, $0.415583$, and $0.593979$, converged to $10^{-9}$, so the finite-difference values carry a grid error of about $10^{-6}$.  The integrated normalized $S_3$ is below $10^{-14}$ (machine precision) in all four tests, consistent with the exact conservation law in Eq.~\eqref{eq:global_S3_zero}.

\end{document}